\documentclass[
aps,
prx,
showpacs,
preprintnumbers,
twocolumn,
superscriptaddress,
10pt,
english,
longbibliography
]{revtex4-2}
\usepackage{hyperref}
\hypersetup{colorlinks,citecolor=blue,linkcolor=blue,urlcolor=blue}

\usepackage{amsfonts}
\usepackage[english]{babel}
\usepackage{graphicx}% Include figure files
\usepackage{dcolumn}% Align table columns on decimal point
\usepackage{bm}% bold math
\usepackage{physics}
\usepackage{xcolor}
\usepackage{mathrsfs}
\usepackage{mathtools}
\usepackage{comment}
\usepackage{multirow} 
\usepackage{array}
\usepackage{enumerate}

\begin{document}
\title{Robust spin-squeezing with random interaction graphs: the lesson from universality}
\author{Andrea Solfanelli}
\email{solfanelli@pks.mpg.de}
\affiliation{Max Planck Institute for the Physics of Complex Systems, Nöthnitzer Str. 38, 01187 Dresden, Germany.}

\author{Augusto Smerzi}
\affiliation{CNR-INO, Largo E. Fermi 6, I-50125 Firenze, Italy.}
\affiliation{LENS, Via N. Carrara 1, Sesto Fiorentino, 50019, Italy.}

%\author{...}
%\affiliation{...}
%\author{...}
%\affiliation{...}

\author{Peter Zoller}
\affiliation{Institute for Quantum Optics and Quantum Information of the Austrian Academy of Sciences, 6020 Innsbruck, Austria.}
\affiliation{Institute for Theoretical Physics, University of Innsbruck, 6020 Innsbruck, Austria.}

\author{Nicol\`o Defenu}
%\email{ndefenu@phys.ethz.ch}
\affiliation{Institut f\"ur Theoretische Physik, ETH Z\"urich, Wolfgang-Pauli-Str. 27 Z\"urich, Switzerland.}
\affiliation{CNR-INO, Area Science Park, Basovizza, I-34149 Trieste, Italy.}

\date{\today} 
\begin{abstract}
We establish the conditions under which scalable spin squeezing can be achieved in interacting spin ensembles embedded in arbitrary, inhomogeneous graph geometries. We identify two different forms of squeezing: OAT-like scalable squeezing is governed solely by the universal properties of the interaction graph and is controlled by its spectral dimension. In critical squeezing, on the other hand, the value of the spectral dimension only furnishes the necessary condition for scalable metrological gain, while the sufficient condition requires the model to lie below the symmetry breaking transition. Therefore, in systems with random interaction graphs, the scaling of the spin-squeezing critical point emerges from a nontrivial interplay between xy-ferromagnetic universality and percolation universality.  We apply this general theoretical framework to several experimental scenarios and discuss sharp and experimentally relevant conditions for achieving robust metrological gain on generic inhomogeneous structures, giving a unifying perspective for designing scalable quantum sensors across diverse quantum simulation platforms.
\end{abstract}

\maketitle
\section{Synopsis}
Quantum-enhanced metrology harnesses many-body entangled states to achieve measurement precision beyond the limits imposed by classical correlations~\cite{giovannetti2011advances,degen2017quantum,pezze2018quantummetrology}. Identifying states suitable for quantum metrology is a delicate challenge: while most states in the Hilbert space exhibit high entanglement, only a few possess the structured correlations necessary for enhanced sensing. Notable examples of metrologically useful quantum states include Greenberger–Horne–Zeilinger states~\cite{bouwmeester1999observation}, Dicke states~\cite{dicke1954coherence}, and squeezed states~\cite{wineland1992spinsqueezing,kitagawa1993squeezed,ma2011quantum}. Designing dynamical protocols which allow for the preparation of metrologically useful states from unentangled product states remains a crucial open problem. These protocols have to be both efficient and robust against noise and disorder in order to apply to a wide range of experimental platforms. The realization of spin squeezing via global interactions has been demonstrated across various platforms, including atomic vapors coupled to light, trapped ions, ultracold gases, and cavity quantum electrodynamics~\cite{pezze2018quantummetrology}.

In this context, the paradigmatic spin-squeezing dynamics is governed by the one-axis-twisting (OAT) Hamiltonian~\cite{kitagawa1993squeezed}
\begin{align}
    H_{\mathrm{oat}} = \frac{S_z^2}{2\mathcal{N}_\mathrm{oat}}\label{eq: OAT Hamiltonian}
\end{align}
which describes a fully connected Ising Hamiltonian and can be interpreted as a planar rotor with moment of inertia determined by the Kac scaling $\mathcal{N}_\mathrm{oat}\sim N$, ensuring energy extensivity~\cite{kac1963van}. Under this assumption, starting from the initial state $\ket{\psi(0)} = \ket{\rightarrow_x}^{\otimes N}$, where the spins are fully polarized in the $x$-direction, and evolving under the OAT Hamiltonian $\ket{\psi(t)} = e^{-iH_{\mathrm{oat}}t}\ket{\psi(0)}$, the optimal spin squeezing in the $yz$-plane perpendicular~($\perp$) to the initial polarization along $x$, is characterized by the squeezing parameter~\cite{wineland1992spinsqueezing,wineland1994squeezed}
\begin{align}
    \xi^2 = \frac{N\min_{\perp}[\mathrm{Var}(S_{\perp})]}{\langle S_x\rangle^2}.\label{eq: squeezing parameter}
\end{align}
which attains its minimum at a time $t_{\min} \sim N^{1/3}$, yielding $\xi^2_{\min} \sim N^{-2/3}$.

The squeezing parameter $\xi^2$ directly controls the phase sensitivity $(\delta\phi)^2 = \xi^2/N$ of a Ramsey-type
measurement~\cite{kitagawa1993squeezed, wineland1992spinsqueezing}. Therefore, a  scalable metrological advantage is achived whenever we obtain scalable spin squeezing, namely a situation in which the minimum in time of the squeezing parameter scales with system size as $\xi^2_{\min} = \xi^2(t_{\min})\sim N^{-\mu}$, with the optimal squeezing reached at a characteristic time scaling as $t_{\min} \sim N^\nu$. Positive scaling exponents $\mu,\nu>0$ then lead to a scaling of the phase sensitivity with the number of sensing spins $(\delta\phi)^2 \sim N^{-(1+\mu)}$ beyond the standard quantum limit~\cite{ma2011quantum}.

In this work we focus on an alternative route to achieve scalable spin squeezing and the associated scalable metrological advantage. In particular we consider a squeezing protocol in which the OAT Hamiltonian~\eqref{eq: OAT Hamiltonian} is replaced by the XXZ model Hamiltonian
\begin{figure*}
    \centering    \includegraphics[width=\linewidth]{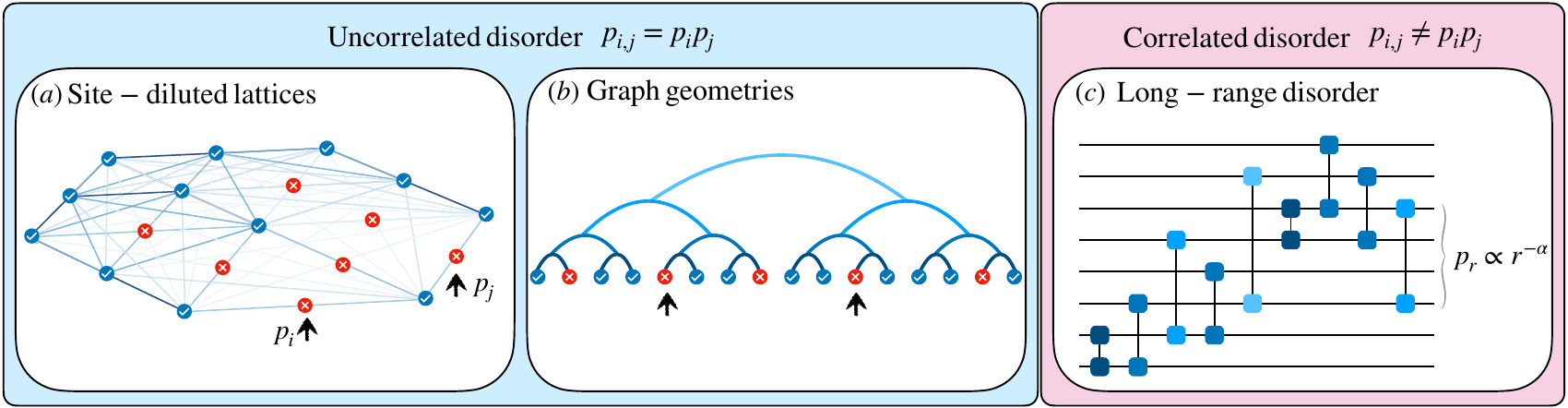}
    \caption{Schematic representations of the three classes of inhomogeneous systems considered in this work: $(a)$ long-range interacting lattices with finite filling fraction or site dilution, modeling realistic experimental conditions in a variety of quantum simulation platforms, including trapped ions, Rydberg atom arrays, polar molecules, and solid-state spin systems such as NV centers (Sec.~\ref{subsec: Diluted long-range lattices}); $(b)$ engineered graph geometries, realizable for instance in Rydberg atom arrays coupled to optical cavities through the application of spatially modulated fields, enabling programmable interaction graphs (Sec.~\ref{subsec: Hypergraphs structures}); $(c)$ systems with spatially correlated disorder, modeling distance-dependent gate errors in digital quantum simulators with limited qubits connectivity or more generally complex spatially correlated imperfections (Sec.~\ref{subsec: The role of spatially correlated disorder}).}   
    \label{fig: Fig1_schematic}
\end{figure*}
\begin{align}
H_{\mathrm{XXZ}} = -\sum_{i,j}J_{ij}\left[s_i^xs_j^x+s_i^ys_j^y+\Delta s_i^zs_j^z\right],\label{eq: G-XXZ Hamiltonian} 
\end{align}
where, $s_i^a$ ($a = x,y,z$) are quantum spin operators with arbitrary spin length $\boldsymbol{s}_i^2 = s(s+1)$. Intuitively, the two squeezing protocols lead to the same metrological advantage for sufficiently connected couplings. In the fully connected case, the XXZ Hamiltonian can be expressed in terms of collective spin operators. Then using the identity $S_x^2+S_y^2 = S^2 - S_z^2$, we recover exactly the OAT model, up to a prefactor $(1-\Delta)$ and an additive constant shift in the Hamiltonian. More precisely, as we will show, the emergence of scalable spin squeezing under the evolution generated by the XXZ Hamiltonian is connected to the development of ferromagnetic order in the $xy$-plane. Accordingly, throughout this work, we restrict the $z$-anisotropy to the regime $|\Delta|<1$. Indeed, outside this regime the system might enter an Ising-like phase with spins polarized along the $z$ direction (ferromagnetically for $\Delta>1$ or antiferromagnetically for $\Delta<-1$), where the mechanism leading to scalable spin squeezing is absent.

Depending on the precise shape of the coupling matrix $J_{i,j}$ and of the $z$-anisotropy parameter $\Delta$, this model can be naturally realized or engineered in diverse quantum simulation platforms, including trapped ions~\cite{britton2012engineered, kiesenhofer2023controlling, guo2024siteresolved,franke2023quantumenhanced}, Rydberg atom arrays~\cite{browaeys2020manybody,gross2017quantum,bornet2023scalable,eckner2023realizing,hines2023spin}, ultracold molecules~\cite{moses2015creation,holland2023ondemand,bilitewski2021dynamical}, and NV centers~\cite{doherty2013nitrogen,gong2023coherent,hughes2025strongly}. 
These physical platforms fall into a large class of many body systems often referred to as long-range interacting systems~\cite{defenu2023long}.
In these systems, the microscopic components interact via a two body coupling matrix which decays as a power-law of their distance $J_{i,j}\propto r_{i,j}^{-\alpha}$. In particular, in trapped-ion qubits, phonon-mediated interactions allow for a tunable exponent in the range $0\lesssim\alpha\lesssim 3$~\cite{britton2012engineered, kiesenhofer2023controlling, guo2024siteresolved}. In Rydberg atoms arrays~\cite{browaeys2020manybody,gross2017quantum}, ultracold molecules~\cite{moses2015creation,holland2023ondemand}, NV centers~\cite{doherty2013nitrogen,gong2023coherent,hughes2025strongly}, and more generically dipolar systems~\cite{chomaz2023dipolar}, the interactions are generated by dipole–dipole couplings, with $\alpha = 3$ or $\alpha = 6$ depending on the dipoles orientation.

The potential for scalable spin squeezing in such systems has recently garnered significant theoretical~\cite{fossfeig2016entanglement,perlin2020spin,comparin2022robust,comparin2022scalble,comparin2022multipartite,roscilde2023entangling,block2024scalable,duha2025nonequilibrium,duha2026twomode} and experimental~\cite{franke2023quantumenhanced, bornet2023scalable,eckner2023realizing,hines2023spin,lee2025observation,douglas2025spin} interest, therefore expanding the class of systems expected to exhibit scalable quantum metrological advantage.

Any realistic experimental implementation, however, is inevitably subject to noise, imperfections, and spatial disorder. Understanding how these imperfections affect spin squeezing is thus essential to understand the conditions under which a robust and scalable quantum metrological advantage can be achieved in realistic experimental scenarios. In this work, we focus on the effects of spatial disorder, motivated by its ubiquity across both digital and analog quantum simulators. 

In particular, spatial disorder naturally arises in several experimental platforms. Recent experiments realizing spin-squeezing dynamics of ensembles of NV centers in a solid-state platform have demonstrated that spatial disorder can strongly degrade squeezing performance, ultimately preventing the achievement of scalable spin squeezing~\cite{wu2025spinsqueezing}. Similarly, spin squeezing in three-dimensional optical lattices has recently been shown to be significantly affected by a finite hole fraction~\cite{lee2025observation}. These findings highlight the need for a deeper theoretical understanding of the role of disorder, both to elucidate its fundamental impact and to devise strategies to overcome its limitations, thereby enabling scalable spin squeezing in experimental platforms relevant for practical quantum sensing applications~\cite{degen2017quantum,maze2008nanoscale,schirhagl2014nitrogen,rovny2024nanoscale,aslam2023quantum}.

Beyond the role played by naturally occurring spatial disorder, the rapid progress of quantum engineering has led to a growing number of experimental platforms capable of achieving quantum many-body dynamics with engineered interaction patterns and controllable disorder. This is the case, for instance, in trapped ions, Rydberg atom arrays, and ultracold quantum gasses in optical lattices. In these programmable systems, spatial disorder is not merely a limitation but can be deliberately introduced and tuned~\cite{sanchezpalencia2010disordered,lewenstein2007ultracold}. In this perspective, the characterization of spin-squeezing dynamics provides a benchmark for the novel physics emerging from the complex geometries which arise from the nontrivial interplay of engineered interactions and disorder. This capability opens the way to direct experimental tests of the theoretical scenarios explored in this work.

All of these platforms are characterized by distinctive forms of  inhomogeneity, which lead to a wide range of different physical situations.

Since our primary goal is to characterize the effects of spatial disorder in experimentally relevant situations, we consider, in general, possibly random graphs where the coupling matrix is drawn from a probability distribution of the form
\begin{align}
\label{bond_prob_dist}
    \Pi[\mathbb{J}_{ij}] = (1-p_{i,j})\delta(\mathbb{J}_{ij}-J_{i,j})+p_{i,j}\delta(\mathbb{J}_{ij}), 
\end{align}
where $p_{i,j}$ denotes the probability that the link between node $i$ and node $j$ is absent. Depending on the choice of $J_{i,j}$ and of $p_{i,j}$ this general framework encompasses a wide variety of experimentally relevant systems within the three classes summarized in Fig.~\ref{fig: Fig1_schematic}.

\textbf{Site-diluted lattices} (Fig.~\ref{fig: Fig1_schematic}a)
describe situations where a spin system is embedded in a translationally invariant lattice, with spins occupying lattice sites. The interaction couplings are ferromagnetic, $J_{i,j}\geq 0$, and decay algebraically with inter-site distance as $J_{i,j}\propto r_{i,j}^{-\alpha}$, with different values of the interaction exponent $\alpha$ corresponding to different physical implementations.

Disorder in these systems arises when a fraction of lattice sites are randomly unoccupied or switched off, producing a diluted random graph (see Fig.~\ref{fig: Fig1_schematic}a). In trapped ions, site dilution can be engineered by stochastically inducing individual ions transitions to atomic states which are effectively decoupled from the dynamics. In neutral atom arrays, ultracold molecules, and solid-state systems, incomplete filling results in a finite probability for each site to be empty~\cite{alonso2010montecarlo,andresen2014existence,kwasigroch2017synchronization,zhang2018quantum,gannarelli2012contribution}. In these cases, the dilution probability is usually assumed to be spatially uncorrelated,  $p_{i,j} = p_ip_j$, and space independent $p_i = p$,  $\forall i$.

\textbf{Graph geometries} (Fig.~\ref{fig: Fig1_schematic}b) can be experimentally realized in neutral atoms arrays within an optical cavity. In this case non-local spin–spin interactions are mediated by photons inside the cavity. The interaction pattern can be finely controlled by applying a magnetic field gradient along the cavity axis and modulating the intensity of the drive field~\cite{periwal2021programmable}. This tunability allows one to program the effective distance dependence of the couplings, thereby engineering geometries whose dimensionality, topology, and metric are entirely distinct from the physical arrangement of the atoms. Remarkably, this approach has enabled the realization of treelike and hypergraph structures inspired by concepts from quantum gravity~\cite{bentsen2019treelike}, in which the interaction graph exhibits sparse but long-range connected topologies. Although the following theoretical formalism applies to any hypergraph and therefore our findings can be applied to a wide range of experimental scenarios, the numerical study will be developed on the illustrative example of the “power-of-two” graph (formally defined in Sec.~\ref{sec: Application to experiments}b), where only nodes separated by distances equal to powers of two are connected.

\textbf{Spatially correlated disorder} (Fig.~\ref{fig: Fig1_schematic}c). In typical experimental situations, including those described in Figs.~\ref{fig: Fig1_schematic}a and~\ref{fig: Fig1_schematic}b, defects or vacancies occur randomly and are uniformly distributed across the system. This scenario is theoretically described by a site dilution with a spatially uniform probability $p$, representing uncorrelated errors that occur independently of the relative positions of the spins. However, depending on the platform or the material under study, more complex situations may arise where imperfections are spatially correlated, leading to strong spatial fluctuations between the experimental defects~\cite{aharonov2006faulttolerant}. The case of spatially correlated disorder is also modeled by the coupling matrix in Eq.~\eqref{bond_prob_dist} with uniform ferromagnetic couplings, $J_{i,j} = J$, combined with a bond dilution probability that depends algebraically on the distance, 
\begin{align}
    p_{i,j} = 1-q_{r}= 1-C r_{i,j}^{-\alpha}.\label{eq: long-range bond probability}
\end{align}
This scenario can be also engineered in digital quantum simulators, where spin–spin couplings are realized through sequences of quantum gates.

In this perspective, the effect of long-range correlated disorder in entanglement scaling and measurement-induced entanglement phase transitions has already been studied~\cite{xu2022longrange,sharma2022measurement,block2022maeasurement}.  

In this work, we characterize the impact of correlated and uncorrelated spatial disorder in quantum metrology. In doing so, we develop a theoretical framework to study the quantum many-body dynamics of interacting spin systems with couplings defined on generic networks~\footnote{Within this work, we use the term \emph{network} in the sense commonly adopted in network theory and complex systems, namely, to denote a graph characterized by nontrivial topological features~\cite{alber2002statistical}. This terminology should not be confused with that of a quantum communication, where a quantum network represents a system of communication links supporting shared entanglement resources.}. The strength and flexibility of our formalism is demonstrated by its application to several experimentally relevant configurations. We demonstrate that the feasibility of scalable spin squeezing is deeply linked to graph universality, particularly the spectral dimension and the possibility of spontaneous symmetry breaking (SSB) of a continuous symmetry in the corresponding inhomogeneous geometry. Our theory demonstrates how different dynamical mechanisms for scalable spin squeezing can be generalized to the case of inhomogeneous and random geometries~\cite{millan2021complex}, paving the way for the realization of useful metrological states on a wide range of quantum platforms.

We now turn to an overview of the main results of the paper. We start by developing a low-energy theory of interacting spin systems on arbitrary graphs, which allows us to identify the universal mechanisms underlying scalable spin squeezing, and to distinguish the two physical routes leading to scalable metrological advantage. Within this framework, the dynamics naturally separates into a collective zero-mode contribution and spin-wave excitations propagating on the graph, whose dynamics is governed by the discrete graph Laplacian. This interplay controls both the emergence of scalable squeezing and its eventual breakdown. Building on these insights, we establish a hierarchy of geometric and dynamical conditions that fully characterize when scalable metrological advantage can be achieved in inhomogeneous quantum systems. Finally we specialize our general results to several experimentally relevant examples.

\subsection{Low energy description and the graph Laplacian}
In order to develop an understanding of the necessary conditions for scalable spin squeezing we first need to develop a low energy model describing the early time dynamics of the system. The low energy excitations of the XXZ model~\eqref{eq: G-XXZ Hamiltonian} are usually described in terms of spin-waves.  However, spin-wave theory is typically carried out in Fourier space taking advantage of the translational invariant properties of the underlying lattice geometry and therefore needs to be generalized to the case of inhomogeneous systems.

Our first result is therefore the introduction of the low energy theory for the XXZ model defined on a generic graph $\mathcal{G}$. In Sec.~\ref{sec: Rotor/Spin-wave theory on graphs} the rotor spin-wave theory method is generalized to the case of the inhomogeneous spin Hamiltonian~\eqref{eq: G-XXZ Hamiltonian}, naturally leading to the emergence of a zero mode, which generates the OAT-like spin squeezing dynamics. On top of this, our treatment demonstrated that spin-wave dynamics is governed by the so called graph Laplacian operator. The graph Laplacian is the discrete analog of the  conventional Laplacian operator $\nabla^2$. Its eigenvalues $\{\lambda_n\}$ play the role of squared momenta in translationally invariant systems $ \lambda_n\approx k^{2}$ on a regular lattice with linear size $L\to\infty$. 
In analogy with the translational invariant case, the relevant time scale governing excitations is then associated to the minimal spectral gap on top of the zero mode: $\delta\lambda = \lambda_1-\lambda_0 = \lambda_1$. Such gap closes in the thermodynamic limit  scaling as $\delta\lambda \sim N^{-2/d_s} \label{eq: spectral gap scaling}$, where $d_s$ is the so called spectral dimension of the graph, a crucial parameter governing the graph universal properties~\cite{cassi1992phase,cassi1996local}, and as we will see also the corresponding spin squeezing dynamics.

\subsection{Physical mechanisms for scalable spin squeezing}
The physical origin of scalable metrological advantage for Ramsey-type sensing experiments in the class of many-body spin systems defined by Eq.~\eqref{eq: G-XXZ Hamiltonian} lies on two alternative mechanisms by which scalable squeezing can be achieved:
\begin{itemize}
    \item[A] \emph{OAT-like spin squeezing}---For sufficiently long-range connected couplings $J_{i,j}$ the XXZ Hamiltonian generates an effective OAT-like dynamics for the collective zero mode, with the moment of inertia set by the graph degree (see Sec.~\ref{sec: Rotor/Spin-wave theory on graphs}). Scalable OAT-type squeezing arises due to the mean-field nature of the ground state transition~\cite{comparin2022robust,comparin2022scalble}.
    
    \item[B] \emph{Critical spin squeezing}---Beyond the mean-field regime scalable spin squeezing can still be achieved in systems that support SSB at finite temperature/energy. Indeed, long-range order translates into collective spin coherence, which in turn unlocks the possibility of scalable spin squeezing~\cite{block2024scalable}. 
\end{itemize}
Long-range interactions induce symmetry breaking also in low-dimenional systems, thus enabling the realization of metrologically useful states based on the previous criteria~\cite{bruno2001absence,maghrebi2017continuous,giachetti2021berezinskii,giachetti2022berezinskii}.
\begin{figure}
    \centering    \includegraphics[width=\linewidth]{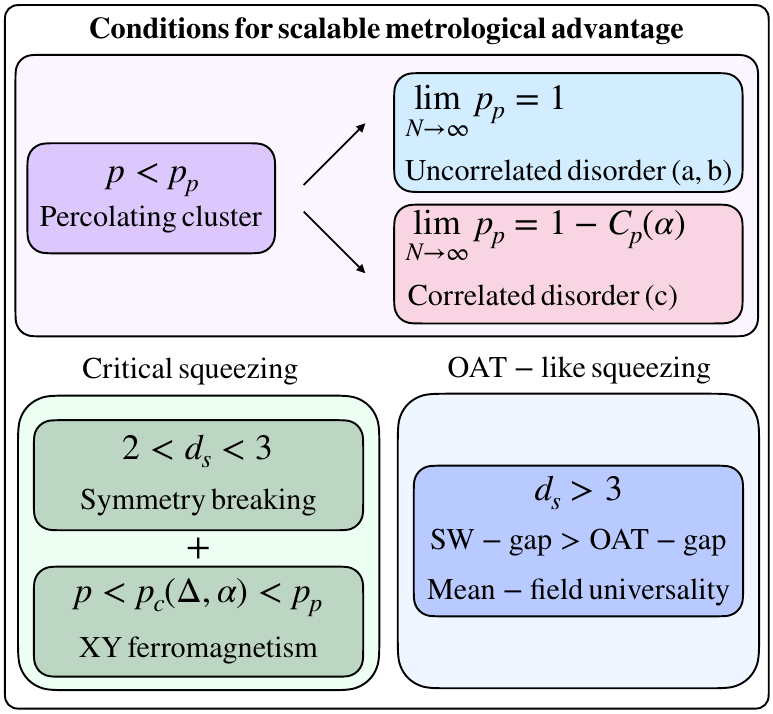}
     \caption{Summary of the hierarchy of necessary and sufficient conditions to obtain scalable metrological advantage on quantum networks. The first, necessary geometric prerequisite is the existence of a giant percolating cluster. On top of this geometric condition, two distinct dynamical mechanisms can lead to scalable squeezing depending on the spectral dimension $d_s$ of the graph: for $d_s>3$, the system lies in the mean-field universality class and exhibits OAT-like squeezing; for $2<d_s<3$, scalable squeezing can still arise due to the existence of the finite temperature transition, see Sec.~\ref{sec: Conditions for scalable spin-squeezing}.}
    \label{fig: Fig_conditionhierarchy}
\end{figure}

\begin{figure*}
    \centering    \includegraphics[width=\linewidth]{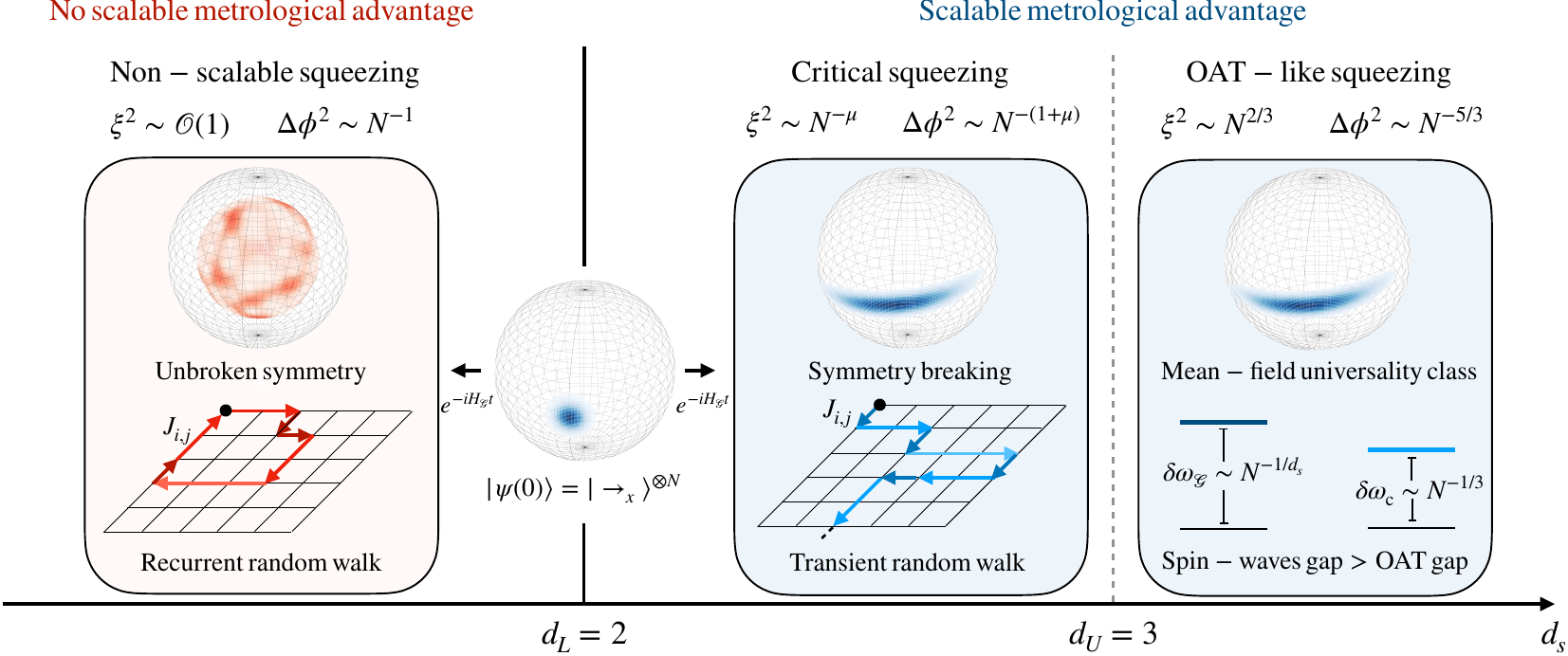}
    \caption{Schematic representation of the necessary conditions for scalable spin squeezing. The possibility of scalable metrological advantage beyond the standard quantum limit is governed by the universality of the spin system on the underlying graph, as encoded by the spectral dimension $d_s$. For $d_s>3$, the system lies in the mean-field regime, where spin-wave excitations are irrelevant on the optimal squeezing timescale, and the dynamics is dominated by the collective zero mode, leading to OAT-like scalable spin squeezing. In the intermediate regime $2<d_s<3$, interactions beyond mean-field become important; nevertheless, the transience of random walks on the graph enables spontaneous symmetry breaking and the establishment of $xy$-ferromagnetic order, giving rise to critical scalable squeezing. In contrast, for $d_s<2$, random walks are recurrent and prevent the formation of long-range order: the system equilibrates to a disordered phase, and no form of scalable spin squeezing can be achieved.}   
    \label{fig: Fig2_SqueezingConditions}
\end{figure*}

\subsection{Conditions for scalable spin squeezing}
\label{sec: Conditions for scalable spin-squeezing}
Before performing any experimental/numerical analysis one must answer the core question:
\emph{Under what conditions on the interaction network $\mathcal{G}$ can scalable spin squeezing and the corresponding metrological advantage be attained? }

Scalable squeezing is a macroscopic property, so any form of it requires that the underlying network $\mathcal{G}$ forms a percolating cluster whose size grows linearly with the system size $\sim N$ in the thermodynamic limit. Once this purely geometric prerequisite is matched further conditions crucially depend on which of the two dynamical mechanisms for squeezing, previously introduced, operates. This leads to two alternative conditions which, as we show in this work, are set apart from the value of the graph spectral dimension $d_s$:
\begin{enumerate}[A]
\item To achieve OAT-type squeezing, the graph Laplacian spectrum must scale at low-energy in order for the OAT-like mode $\mathbf{u}_{0}$ dynamics to occur on a different timescale from that of the high-energy modes $\mathbf{u}_{n>0}$. This condition is satisfied for values of the spectral dimension such that $d_s>3$ so that the quantum critical point governing the low energy dynamics lies in the mean-field universality class. 
\item Critical squeezing needs the underlying network to support SSB of a continuous symmetry, which is governed by the recurrence properties of the random walker in the network $\mathcal{G}$. 
\end{enumerate}
For OAT-type spin squeezing having a percolating cluster and condition A constitute a set of necessary and sufficient conditions. On the other hand for the case of critical squeezing conditions B is only necessary as it only establishes that an ordered phase may exist. Accordingly we need to specify the critical values of the microscopic parameters at which such a phase is actually realized. 
Conditions A and
B apply to complementary regimes of $d_s$ and are not alternatives. The complete hierarchy of conditions needed for scalable spin squeezing in the different regimes are summarized Fig.~\ref{fig: Fig_conditionhierarchy}.
Let us analyze these conditions one by one.
\subsubsection{Geometric prerequisite: percolation}
The first, purely geometric requirement is the existence in the graph $\mathcal{G}$ of a giant percolating cluster, which is necessary for any collective behavior to emerge. This condition already sets aside uncorrelated and correlated forms of disorder. 

In the case of uncorrelated site dilution, where each site is removed independently with probability $p$,  the critical percolation threshold $p_p$ depends on the connectivity of the underlying clean graph. In particular, as shown in Sec.~\ref{subsec:Spatially uncorrelated disorder}, due to the all-to-all connectivity of the underlying long-range interacting lattice we have that $\lim_{N\to\infty}p_p\to 1$. As a consequence, in an infinite site diluted long-range interacting graph, the existence of a giant percolating cluster is guaranteed for any finite dilution $p<1$. 

Nevertheless, as we will show in the concrete examples discussed in Secs.~\ref{subsec: Diluted long-range lattices} and~\ref{subsec: Hypergraphs structures}, the condition~\eqref{eq: critical p} remains relevant for finite-size experiments with limited filling fractions $f = 1-p$, such as ultracold polar molecules in optical lattices ($p\sim 0.8$)~\cite{moses2015creation} or ensembles of solid-state NV centers ($p\sim 0.999$) ~\cite{gong2023coherent,hughes2025strongly}. In practice, for any finite $N$ there exists a critical dilution probability (or equivalently a critical filling fraction) that marks the onset of scalable spin squeezing.

The presence of a spatially correlated disorder drastically changes the picture, as detailed in Sec.~\ref{subsec: The role of spatially correlated disorder}. There, a finite percolation threshold can persist even as $N\to\infty$, imposing a fundamental limit on scalable squeezing. 

\subsubsection{Spectral gap criterion for OAT-type squeezing}
The presence of a percolating cluster, while necessary, is not sufficient for scalable spin squeezing. 

In order to identify the additional requirements for OAT-like spin squeezing on a network geometry we first need to understand the universal origin of the $t_{\mathrm{min}}\sim N^{1/3}$ scaling of the time to achieve the minimum of the squeezing parameter. Since the initial state is polarized along the 
$x$-direction, the OAT dynamics can be naturally interpreted as a quench from the ground state of a fully connected quantum Ising model in a transverse field, also known as the Lipkin–Meshkov–Glick (LMG) model \cite{glick1965validity}, from the paramagnetic phase to the ferromagnetic phase at zero field, thereby driving the system across its quantum critical point. Within this picture, the optimal squeezing time is controlled by the finite-size scaling of the energy gap between the ground state and the first excited state at criticality, which sets the relevant dynamical timescale of the evolution. Specifically, the critical gap scales as $\delta\omega_{\mathrm{c}}\sim N^{-z}$, where $z= 1/3$ is the dynamical critical exponent for the LMG model~\cite{botet1982size,dusuel2004finite}. Indeed, $\delta\omega_c$ determines the time scale governing the dynamics of the lowest energy modes, which are responsible for scalable spin squeezing~\footnote{Notice that spin-squeezing protocols with different scaling in time can be designed \cite{micheli2003manyparticle,roscilde2025exponential} but in this case the quench crosses different transition lines \cite{ ribeiro2008exact,munozarias2023phasespace}}.

This observation naturally leads to the following requirement for the spectral gap $\delta\omega_\mathcal{G}$ in a network, which must be compared with $\delta\omega_\mathrm{c}$:
\begin{align}
    \delta\omega^{-1}_\mathcal{G}\sim N^{1/d_s}
    <\delta\omega^{-1}_\mathrm{c}\sim t_{\mathrm{min}}\sim N^{1/3}.\label{spectral_gap_condition}
\end{align}
If this condition is met, the dynamics of the low-energy modes evolve rapidly enough to achieve optimal spin squeezing as in the OAT model. Interestingly, the spectral gap $\delta\omega_\mathcal{G}$, which, as shown in Sec.~\ref{sec: Rotor/Spin-wave theory on graphs}, for $\Delta\neq 1$ corresponds to the square root of the of the graph Laplacian spectral gap $\delta\lambda$, which is also known as the graph algebraic connectivity~\cite{fiedler1973algebraic}. The spectral gap is also responsible for fast information propagation~\cite{millan2021local,chung1997spectral} and optimal spatial search times in quantum walks on random graphs and lattices~\cite{childs2004spatial,chakraborty2020optimality, king2025optimal}.

Equation~\eqref{spectral_gap_condition} translates into a condition on the graph spectral dimension $d_s>3$, indicating that the effective dimension governing universality must exceed the upper critical dimension $d_u$ of the corresponding nearest-neighbor interacting quantum spin model (see Fig.~\ref{fig: Fig2_SqueezingConditions}). This condition generalizes the known criterion for spin squeezing in the XY model with power-law decaying interactions $J_r\propto r^{-\alpha}$ to generic interaction networks. Specifically, the spectral dimension for the XY model in long-range lattices relates to $\alpha$ through $d_s = 2d/(\alpha-d)$ in the mean-field regime~\cite{solfanelli2024universality}. 

Given the fully-connected nature of the OAT Hamiltonian, Eq.~\eqref{spectral_gap_condition} is equivalent to require that the quantum critical point of the XXZ Hamiltonian on the graph lies in the mean-field universality class
\begin{align}
\label{duc_condition}
d_{s}>d_{u}\Rightarrow \alpha<5d/3,
\end{align}
where $d_{u}=3$ is the upper critical dimension of the quantum XXZ model. The r.h.s. of Eq.~\eqref{duc_condition} is consistent with the result on regular lattices~\cite{comparin2022robust}.
\subsubsection{Critical squeezing and random-walk transience}
When $d_s < d_u$, spin squeezing may still occur, but it is no longer governed solely by the low-energy properties of the model. In this regime, the rotor/spin-wave approximation becomes insufficient to fully characterize the dynamics at the time scales where the $N$-dependent minimum of $\xi^2$ is reached. Indeed, for $\Delta \neq 1$, the initial state has finite overlap with eigenstates across a range of finite energy densities. There, the squeezing dynamics occurs due to the critical states responsible for the finite-temperature transition.

Therefore, spin squeezing can emerge at $d_{s}<d_{u}$ when the energy density of the initial state lies below the critical temperature associated with the onset of $xy$-ferromagnetism. Thus, the necessary condition to achieve a squeezed state on an inhomogeneous network with $d_{s}<d_{u}$ coincides with the possibility for SSB of the continuous symmetry to occur. In non-homogeneous structures, SSB is fundamentally linked with the recurrence properties of random walks on the corresponding network geometry~\cite{cassi1992phase}.
 
In graphs with spectral dimension $d_s \leq 2$, random walks are recurrent: the walker returns to its origin with probability one, preventing the establishment of long-range order~\cite{cassi1992phase}. For $d_s>2$, random walks typically become transient allowing for the establishment of a macroscopic order parameter~\cite{burioni1999transience} and thus the occurrence of scalable spin squeezing (see Fig.~\ref{fig: Fig2_SqueezingConditions}).

Together with Eq.~\eqref{duc_condition}, the condition $d_{s}>2$~\cite{burioni1999inverse,burioni2005random}, yields the finite window 
\begin{align}
\label{critical_squeezing_window}
2<d_{s}<3
\end{align}
for the occurrence of critical spin squeezing (see Fig.~\ref{fig: Fig2_SqueezingConditions}). However, this condition, similarly to the one that implies the existence of a percolating cluster, is only a necessary condition. For critical spin squeezing to occur, one must also ensure that the initial state energy lies below the scale set by the critical temperature of the Hamiltonian~\eqref{eq: G-XXZ Hamiltonian}. 

\subsubsection{Scaling of the critical point in proximity to the percolation threshold}\label{subsec: Scaling of the critical point in proximity to the percolation threshold}
\begin{table*}
    \centering
    \renewcommand{\arraystretch}{1.4} % row height
    \setlength{\tabcolsep}{2pt} % column spacing
    \begin{tabular}{|p{3.5cm}|p{3.5cm}|p{5cm}|p{5cm}|}
         \hline
         \centering Type of inhomogeneity 
         & \centering Interaction shape \arraybackslash
         & \centering Scalable spin squeezing conditions \arraybackslash
         & \centering Experimental platforms \arraybackslash \\
         \hline
         
         \multirow{4}{=}{\centering Diluted long-range site lattices~\ref{subsec: Diluted long-range lattices}} 
         &\multirow{2}{=}{\centering $J_r \propto 1/r^{\alpha}$} 
         &\multirow{1}{=}{$\alpha < 2d$} &Trapped ions ($1\lesssim\alpha\lesssim 3$)~\cite{britton2012engineered, kiesenhofer2023controlling, guo2024siteresolved}\\
         & &\multirow{1}{=}{$p <p_c(\Delta,N)<p_\mathrm{p}(N)$} &Rydberg Atoms ($\alpha=3$)~\cite{browaeys2020manybody,gross2017quantum}\\
         &\multirow{2}{=}{\centering $p_{\mathrm{site}} = \mathrm{const.}$} &\multirow{1}{=}{$p_\mathrm{p}(N) = 1-(N-2)^{-1}$}  &Molecules $(\alpha=3$)~\cite{moses2015creation,holland2023ondemand}\\
         & &\multirow{1}{=}{$p_p(N)-p_c(\Delta,N)\approx(1-\Delta)^{d/(\alpha-d)}$} & NV centers ($\alpha=3$)~\cite{doherty2013nitrogen,gong2023coherent,hughes2025strongly}\\
        \hline
        \multirow{4}{=}{\centering Graph geometries~\ref{subsec: Hypergraphs structures}} &\multirow{4}{=}{\centering $J_r\propto\begin{cases}
         r^{-\alpha} & r = 2^{n}\\
        0&\mathrm{otherwise}\end{cases}$}  &\multirow{1}{=}{$\alpha< 1$} &\multirow{4}{=}{Cold atoms in optical cavities~\cite{periwal2021programmable}}\\ 
        & &\multirow{1}{=}{$p <p_c(\Delta,N)<p_\mathrm{p}(N)$} &\\
        & &\multirow{1}{=}{$p_p(N) = 1-[\log_2(N)-5/2-1/N]^{-1} $} &
        \\
        & &\multirow{1}{=}{$p_p(N)-p_c(\Delta,N)\approx(1-\Delta)^{1/\alpha}$} &
        \\
        \hline
          \multirow{4}{=}{\centering Correlated disorder~\ref{subsec: The role of spatially correlated disorder}} &\multirow{2}{=}{\centering$p_\mathrm{bond}(r) = 1-C/r^{\alpha}$} &\multirow{1}{=}{$\alpha<2$} &\multirow{4}{=}{Gate based quantum processors~\cite{cheng2023noisy,weaving2023benchmarking, solfanelli2024stabilization}}
          \\
          & &\multirow{1}{=}{$C>C_c(\Delta,\alpha)\geq C_p(\alpha)$} &\\
          &\multirow{2}{=}{\centering$J_r = \mathrm{const.}$} &\multirow{1}{=}{$ C_p(\alpha)>1-1/2\zeta(\alpha)$} &\\
          & &\multirow{1}{=}{$ C_p(\alpha)-C_c(\Delta,\alpha)\approx (1-\Delta)^{d_s/2\gamma}$} &\\
         \hline
    \end{tabular}
    \caption{Overview of the classes of inhomogeneous spin systems investigated in this work. For each class, we report the type of spatial inhomogeneity, the functional form of the interactions, and the corresponding conditions under which scalable spin squeezing is achieved. These conditions are expressed in terms of the interaction exponent $\alpha$, the anisotropy $\Delta$, and the relevant percolation or connectivity thresholds, highlighting the role of both disorder and graph topology. The last column indicates representative experimental platforms where the corresponding interaction patterns and dilution mechanisms naturally arise.}
    \label{tab: Summary application to experiments}
\end{table*}
The relation between the possibility of spontaneous symmetry breaking and the transience of random walks on a graph is rooted in rigorous results that do not rely on low-energy assumptions~\cite{cassi1992phase}. However, the condition $d_s>2$ only establishes that an ordered phase may exist; it does not specify the critical values of the microscopic parameters at which such a phase is actually realized.

In the region $2<d_{s}<3$, the finite temperature phase transition lies outside the mean-field regime and strong fluctuations hinder reliable estimation of the critical quantities, including the location of the critical point. However, in order to build a theoretical expectation for the critical value of the $z$-anisotropy $\Delta$ in our model and benchmark the following numerical simulations (see Sec.~\ref{sec: Application to experiments}), we focus on the $\Delta \to 1$ limit, where a perturbative approach can be constructed. There, the spin-wave Hamiltonian reduces exactly to the graph Laplacian (see Sec.~\ref{sec: Rotor/Spin-wave theory on graphs}), and the dynamics of single-particle excitations is exactly governed by diffusion on the network. Deviations from this point introduce an anisotropy term proportional to $1-\Delta$ , which can be treated as a perturbation, see Sec.~\ref{subsec: Perturbation theory close to the Heisenberg point at strong dilution} for additional details.

The stability of the ordered phase is then determined by the competition between the energy scale associated with this perturbation $\delta\varepsilon_\Delta\propto(1-\Delta)$ and the spectral gap of the unperturbed Hamiltonian, given by the graph algebraic connectivity $\delta\lambda$. Estimating the matrix element of the perturbation between the ground state and the first excited state (see~\ref{subsec: Perturbation theory close to the Heisenberg point at strong dilution} for the derivation) yields the criterion
\begin{align}
\delta\varepsilon_{\Delta_c}\propto(1-\Delta_c)\sim\delta\lambda\label{eq: delta_eps-delta_lamb} 
\end{align}
In clean lattices, the Laplacian spectrum is gapless in the thermodynamic limit, $\delta\lambda\to 0$, implying that the anisotropy is always a relevant perturbation. As a result, the precise value of 
$\Delta_c$ and hence the critical temperature for $xy$-ordering depends nonuniversally on microscopic details of the model. However, a universal scaling of $1-\Delta_c$ as a function of $|p_p-p|$, can still be achieved in the presence of strong dilution when the system approaches the percolation transition.

Indeed, the scaling of the spectral gap with system size~\eqref{eq: spectral gap scaling} needs to be modified close to the percolation threshold by replacing the number of nodes $N$ with the average size of the percolating cluster $S$, Eq.~\eqref{eq: delta_eps-delta_lamb} then leads to 
\begin{align}
    1-\Delta_c\sim S^{-2/d_S}.
\end{align}
The scaling of $S$ close to the percolation transition $p\to p_p$ is governed by the $\gamma$ critical exponent~\cite{benAvraham2000diffusion}
\begin{align}
    S\sim |p_p-p|^{-\gamma},
\end{align}
Combining these results yields the scaling behavior 
\begin{align}
   1-\Delta_c\sim |p_p-p|^{2\gamma/d_s}.\label{eq: Delta_c scaling}
\end{align}
which describes how the critical anisotropy approaches the Heisenberg point as the percolation threshold is approached.

This behavior has a clear physical interpretation: near the percolation transition, the suppression of connectivity weakens the stability of the ordered phase, requiring increasingly isotropic interactions ($\Delta\to 1$) to keep the energy of the $x$-polarized state below the critical value for $xy$-ferromagnetism. Indeed, the $x$-polarized state becomes a ground state at $\Delta = 1$, hindering the realization of scalable spin squeezing (see Sec.~\ref{subsec: The role of spatially correlated disorder} for a direct comparison of the $\Delta \to 1$ and $p\to p_p$ limit of the critical point with the prediction in Eq.~\eqref{eq: Delta_c scaling}). 
\subsection{Summary of the main results for different types of inhomogeneity}
Several experimental platforms in which different forms of inhomogeneity can be realized are discussed in Sec.~\ref{sec: Application to experiments} and summarized in the last column of Tab.~\ref{tab: Summary application to experiments}. As a concrete example, in trapped-ion systems the full phase diagram of scalable spin squeezing can be explored by experimentally tuning three key parameters:
\begin{enumerate}
    \item The interaction exponent $\alpha$ of the power-law couplings $J_r \propto r^{-\alpha}$, which directly controls the spectral dimension $d_s$ of the effective interaction graph~\cite{millan2021complex, bighin2024universal, solfanelli2024universality}. Experimentally, $\alpha$ can be tuned by adjusting the trapping frequencies and the and the detuning between the driving lasers and the ions motional modes~\cite{britton2012engineered, kiesenhofer2023controlling, guo2024siteresolved}.
    \item The dilution probability $p$, which can be engineered by selectively addressing individual ions and transferring them to auxiliary internal states that are effectively decoupled from the dynamics.
    \item The $z$-anisotropy $\Delta$ of the XXZ Hamiltonian~\eqref{eq: G-XXZ Hamiltonian}, which can be controlled using Floquet engineering techniques~\cite{kranzl2023observation,bukov2015universal}.
\end{enumerate}

A central outcome of our analysis is that spatially uncorrelated disorder, such as site dilution, does not affect the universality of the squeezing dynamics. In particular, the spectral dimension $d_s$ remains independent of the dilution probability $p$. Moreover, in diluted geometries where the connectivity of the clean graph scales with the system size, no true percolation transition occurs in the thermodynamic limit ($N\to\infty$), implying that scalable spin squeezing is robust against finite filling fractions, provided the system remains within the $xy$-ordered phase (see Sec.~\ref{subsec: Diluted long-range lattices} and ~\ref{subsec: Hypergraphs structures}).

Finite-size effects, however, can still play an important role in realistic systems. A qualitative distinction emerges between lattice-based and more general graph-based geometries. While both can share the same spectral dimension, the scaling of finite-size corrections differs significantly. In particular, for the PW2 graph example, considered in Section~\ref{subsec: Hypergraphs structures}, corrections decay only logarithmically with system size, $\sim 1/\log_2(N)$, as opposed to the faster $\sim 1/N$ scaling typical of regular lattices. As a consequence, finite-size effects are substantially more pronounced in sparse or hierarchical graph structures, leading to a stronger suppression of squeezing in experimentally accessible system sizes.

Finally, spatially correlated disorder, modeled through distance-dependent bond activation probabilities, leads to qualitatively new behavior (see Sec.~\ref{subsec: The role of spatially correlated disorder}). In this case, a finite percolation threshold persists even in the thermodynamic limit, fundamentally constraining the emergence of long-range order. The resulting phase diagram is governed by a nontrivial interplay between percolation criticality and $xy$-ferromagnetic universality. This interplay gives rise to a universal scaling of the critical anisotropy, as captured by Eq.~\eqref{eq: Delta_c scaling}, where critical exponents from both percolation and magnetic transitions enter explicitly.

\section{Generalized Rotor/Spin-wave theory on graphs}\label{sec: Rotor/Spin-wave theory on graphs}
In this Section we introduce the generalized rotor/spin-wave theory framework to describe the low energy theory of spin systems interacting  on generic network structures. Additional details on the results obtained for the spin squeezing dynamics within this setup and the comparison of the approximation with a more standard linear spin wave theory can be found in App.~\ref{app: Rotor-Spin wave}.

\subsection{Reminders of graph theory}
Let us start by briefly sumarizing the graph theory concepts needed to develop the generalize low energy description of the XXZ model~\eqref{eq: G-XXZ Hamiltonian} to the case in which the underlying couplings geometry is a generic graph.  

We define a graph $\mathcal{G}$  as a countable set $V$ of vertices (or nodes) $i$ connected pairwise by an unoriented edge (or bonds) set $E$ with links $(i,j) = (j,i)$. The topology of the graph is encoded in its adjacency matrix $\mathbb{A}$, defined as
\begin{align}
    \mathbb{A}_{ij} = \begin{cases}
        1 &\mathrm{if}\,(i,j) \in E\\
        0 &\mathrm{otherwise}
    \end{cases}.
\end{align} 
The coupling matrix $\mathbb{J}$ defines the interaction strengths:
\begin{align}
    \mathbb{J}_{ij} = \mathbb{J}_{ji} = \begin{cases}
        J_{ij}>0 &\mathrm{if}\quad\mathbb{A}_{ij} = 1\\
        0 &\mathrm{if}\quad\mathbb{A}_{ij} = 0
    \end{cases}.\label{eq: coupling matrix}
\end{align}
We also define the degree matrix $\mathbb{D}$, a diagonal matrix whose entries are given by the weighted coordination numbers 
\begin{align}
\mathbb{D}_{jj} = \sum_{l} J_{lj}.\label{eq: degree matrix}
\end{align} 

A central object for characterizing the graph spectral properties, which are strongly related to the system universality, is the generalized graph Laplacian which is defined as 
\begin{align}
    \mathbb{L} = \mathbb{D} - \mathbb{J}\label{eq: graph Laplacian}
\end{align}
Being Hermitian, $\mathbb{L}$ can be diagonalized, yielding a spectrum of eigenvalues ${\lambda_n}\geq 0$ and the corresponding orthonormal eigenvectors ${\boldsymbol{u}_n}$ satisfying $\mathbb{L} \boldsymbol{u}_n = \lambda_n \boldsymbol{u}_n$. We order the eigenvalues as $\lambda_0\leq\lambda_1\leq\dots\lambda_{N-1}$, with $\lambda_0 = 0$ and the corresponding eigenvector $\boldsymbol{u}_0 = (1, 1, \dots, 1)$, representing the uniform mode.

To gain further insight into the system spectral properties, it is useful to introduce the Laplacian density of states
\begin{align}
    \rho(\lambda) = \frac{1}{N}\sum_{n}\delta(\lambda-\lambda_n).
\end{align}
A key spectral feature is the spectral gap $\delta\lambda = \lambda_1 - \lambda_0 = \lambda_1$, also known as the algebraic connectivity of the graph~\cite{fiedler1973algebraic}. The scaling of this gap with system size is governed by the spectral dimension $d_s$, which is defined by the small $\lambda$ behavior of $\rho(\lambda)$: \begin{align} 
\rho(\lambda) \sim \lambda^{d_s/2 - 1} \quad \text{as } \lambda \to 0, \end{align} 
leading to the gap scaling relation in Eq.~\eqref{eq: spectral gap scaling}.

The spectral dimension $d_s$ governs the diffusion on the graph. In particular, the dynamics of a random walk on the weighted graph $\mathcal{G}$ is defined by the transition matrix $(\mathbb{D}^{-1}\mathbb{J})_{ij} = J_{ij}/(\sum_kJ_{ik}$), where $\mathbb{J}$ is the coupling matrix~\eqref{eq: coupling matrix} and $\mathbb{D}$ the degree matrix~\eqref{eq: degree matrix}. Thus, the random work hops between the two sites $i,j$ with transition amplitude $(\mathbb{D}^{-1}\mathbb{J})_{ij}$ and has a probability $P_{ij}(t)$ of reaching any site $j$ at time $t$ after having initiated its dynamics at site $i$ with $t=0$. The random walk dynamics obeys the master equation~\cite{burioni2005random}
\begin{align}
    \frac{dP_{i,j}(t)}{dt}&= -\sum_{k}(\mathbb{I}-\mathbb{D}^{-1} \mathbb{J})_{i,k}P_{k,j}(t)\notag\\
    &=-\sum_{k}(\mathbb{D}^{-1}\mathbb{L})_{i,k}P_{k,j}(t)
\end{align}
where $\mathbb{L}$ is the graph Laplacian~\eqref{eq: graph Laplacian}. Then, the Laplacian not only governs the spin-wave spectrum but also controls diffusion processes on the network and the long time behavior of the recurrence probability~\cite{burioni2005random} is determined by the spectral dimension as
\begin{align}
   \langle P(t)\rangle_\mathcal{G}= \frac{1}{N}\sum_i P_{ii}(t)\sim t^{-d_{s}/2}\,.\label{eq: recurrence probability ds}
\end{align}
In Sec.~\ref{sec: Application to experiments} we will use Eq.~\eqref{eq: recurrence probability ds} as a further benchmark for the numerical calculation of $d_{s}$. 

\subsection{Linear spin-wave theory}
The starting point is the mapping of spins to bosons via the Holstein–Primakoff (HP) transformation~\cite{holstein1940field}, with the quantization axis chosen along the $x$-direction
\begin{subequations}
\label{hp_eq}
\begin{align}
    s^x_l &= s-a_l^\dagger a_l\\
    s^y_l &= (\sqrt{2s-n_l}a_l+\mathrm{h.c.})/2,\\
    s^z_l &= (\sqrt{2s-n_l}a_l-\mathrm{h.c.})/2i,
\end{align}
\end{subequations}
where $a_l$ and $a_l^\dagger$ are bosonic operators, $n_l=a_l^\dagger a_l$. If the Hamiltonian in Eq.~\eqref{eq: G-XXZ Hamiltonian} exhibits ferromagnetic long-range order in the $xy$-plane, one can choose the mean-field ground state to be a coherent spin state with all spins polarized along the $x$-axis, $\ket{\mathrm{CSS}_x} = \ket{\to_x}^{\otimes N}$. This state acts as the vacuum of the HP bosons defined in Eq.~\eqref{hp_eq}. Expanding the HP transformation at leading order in $s$, the Hamiltonian~\eqref{eq: G-XXZ Hamiltonian} takes the form
\begin{align}
    H_{\mathrm{\mathcal{G}-XXZ}}= E_{mf}+H_2+\dots
\end{align}
where $E_{\mathrm{mf}} = -\sum_{i,j} J_{i,j} s^2$ is the mean-field energy and $H_2$ describes the quadratic fluctuations around it. Introducing the vector of bosonic operators $\Psi^\dagger = (a^\dagger_1,\dots, a^\dagger_N, a_1,\dots, a_N)$,  the quadratic Hamiltonian can be rewritten as  $H_{2} = \Psi^{\dagger}\mathbb{H}\Psi$, where $\mathbb{H}$ is a $2N\times 2N$ matrix given by
\begin{align}
    \mathbb{H} = \begin{pmatrix}
        s[\mathbb{D}-\mathbb{J}(1+\Delta)/2] &-s\mathbb{J}(1-\Delta)/2\\
        -s\mathbb{J}(1-\Delta)/2 &s[\mathbb{D}-\mathbb{J}(1+\Delta)/2]
    \end{pmatrix},\label{eq: sw Hamiltonian matrix}
\end{align}
where $\mathbb{J}$ is the coupling matrix~\eqref{eq: coupling matrix} and $\mathbb{D}$ the diagonal degree matrix~\eqref{eq: degree matrix}.

The key to tackle the case of inhomogeneous non-translational invariant geometries is to introduce a generalized rotor/spin-wave theory in which we decompose the bosonic Hamiltonian in terms of the eigenmodes of the graph Laplacian operator $\mathbb{L}$ defined in Eq.~\eqref{eq: graph Laplacian}. The dispersion relation of the spin-waves is mapped to the Laplacian spectrum $\{\lambda_n\}$, and the density of states $\rho(\lambda)$ encodes the effective phase space available to low-energy excitations. The graph Laplacian zero mode, whose eigenvector $\mathbf{u}_{0}$ is uniformly distributed over all the graph nodes, plays the role of the  rotor and generates the OAT-like dynamics. The other Laplacian eigenmodes  $\mathbf{u}_{n>0}$ describe the bosonic excitations that spread throughout the network.

\begin{figure*}
    \centering    \includegraphics[width=\linewidth]{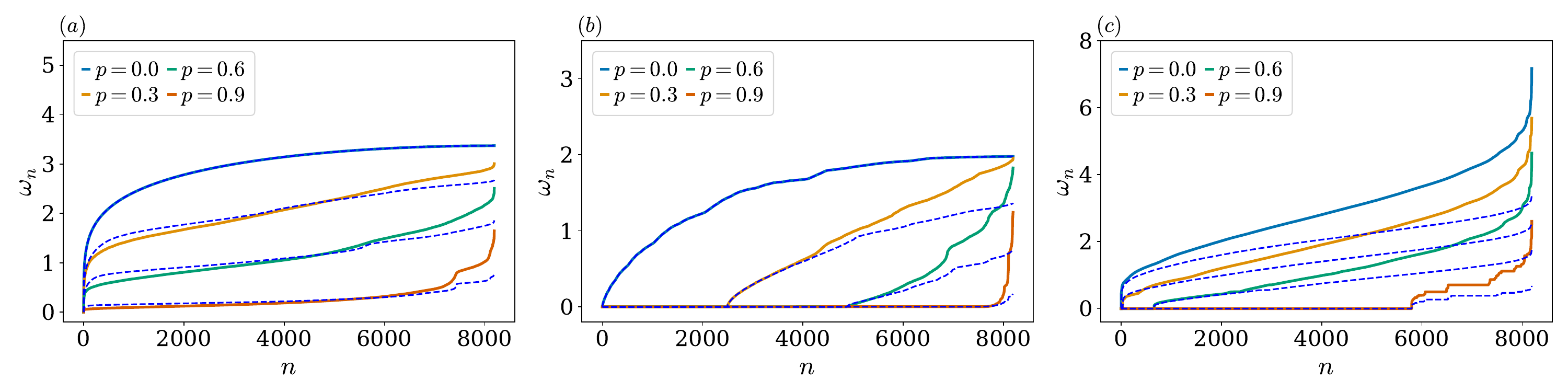}
    \caption{Comparison of the spin-wave spectra obtained by exact diagonalization of the real-space quadratic Hamiltonian~\eqref{eq: sw Hamiltonian matrix} (dots), and by the regular graph approximation~\eqref{eq: average degree spectrum} (blue dashed lines), for different dilution probabilities. (a) One-dimensional site-diluted long-range lattice with power-law interactions, system size $N = 8192$  and interaction exponent $\alpha = 1.4$. (b) PW2 graph with $N = 8192$ and $\alpha = 1.4$. (c) Graph with power-law spatially correlated disorder, where the bond are deleted with probability $p_{i,j} =1- C/r_{i,j}^\alpha$ shown for $C = 1-p$ and $\alpha = 1.4$.}
   \label{fig: SpinWave_Spectrum}
\end{figure*}
Thus, for the XXZ case with $\Delta\neq 1$, the spin waves Hamiltonian does not reduce to the graph Laplacian. However, if the underlying interaction graph is regular, i.e., all nodes share the same degree, $\sum_{l}J_{l,j} = \mathrm{deg}_\mathcal{G}$ $\forall j$, then $\mathbb{D} = \mathrm{deg}_\mathcal{G}\mathbb{I}$ and the coupling matrix $\mathbb{J}$ can be diagonalized in the same basis of the graph Laplacian. Then, the eigenvalues $\{J_n\}$ of $\mathbb{J}$ are related to those of the Laplacian via $J_n =\mathrm{deg}_\mathcal{G}-\lambda_n $, where $\lambda_0 = 0$ corresponds to the uniform eigenvector $\mathbf{u}_0 = (1,1\dots,1)$. 

Within the hypothesis of a regular graph, the bosonic operators can be represented in the eigenbasis of the Laplacian $\{\mathbf{u}_n(r_j)\}$, analogously to the plane-wave expansion in translationally invariant systems, leading to  
\begin{align}
    a_j = \frac{1}{\sqrt{N}}\sum_n\mathbf{u}_n(\mathbf{r}_j)a_n\,,
\end{align}
where the normalization conditions $\sum_j\mathbf{u}_n(\mathbf{r}_j)\mathbf{u}_m(\mathbf{r}_j) = N\delta_{n,m}$ and $\sum_{j}\mathbf{u}_n(\mathbf{r}_j)\mathbf{u}_n(\mathbf{r}_l) = N\delta_{j,l}$ have been assumed. Rewriting $H_2$ in terms of the $\Psi_n^\dagger = (a_n^\dagger,a_n)$ operators, we obtain $H_2 = \sum_{n}\Psi_n^\dagger\mathcal{H}_n\Psi_n$, where each $2\times 2$ block $\mathcal{H}_n$ is given by
\begin{align}
    \mathcal{H}_n = \begin{pmatrix}
        s[\mathrm{deg}_\mathcal{G}-J_n(1+\Delta)/2] &-sJ_n(1-\Delta)/2\\
        -sJ_n(1-\Delta)/2 &s[\mathrm{deg}_\mathcal{G}-J_n(1+\Delta)/2]
    \end{pmatrix},
\end{align}
and $J_n = \sum_{i,j}\mathbf{u}_n(\mathbf{r_i})J_{i,j}\mathbf{u}_n(\mathbf{r_j})$ are the eigenvalues of the coupling matrix.

The quadratic Hamiltonian is diagonalized via a Bogoliubov transformation introducing bosonic operators $\alpha_n$, $\alpha_n^\dagger$ such that $a_n = \mathcal{U}_n\alpha_n-\mathcal{V}_n\alpha_n^\dagger$, 
yielding the diagonal form $H_2 = \sum_n\omega_n\alpha_n^\dagger\alpha_n$, where the spin-wave dispersion relation can be written in terms of the graph Laplacian eigenvalues as
\begin{align}
    \omega_n
    = s\sqrt{\lambda_n(\Delta\lambda_n+\mathrm{deg}_\mathcal{G}(1-\Delta))}.
\end{align} 
As in the translational invariant case, the zero mode excitation energy vanishes ($\omega_0=0$) and the spectral gap is controlled by the Laplacian gap $\delta\lambda$:
\begin{align}
\delta\omega_\mathcal{G}\approx\begin{cases}
    \sqrt{\delta\lambda J_0(1-\Delta)}&\Delta\neq 1\\
    |\delta\lambda| &\Delta = 1
\end{cases}.
\end{align}
\subsection{The zero mode contribution}
Due to the vanishing zero-mode energy, the Bogoliubov transformation becomes singular at $n = 0$, indicating that the population of the $n = 0$ bosons $\langle a_0^\dagger a_0\rangle$ cannot be considered as a small perturbation and therefore calling for a separate treatment of the nonlinear terms involving the $a_0$ , $a_0^\dagger$ (see App.~\ref{app: Rotor-Spin wave} for the details on how standard linear spin-wave theory fails in correctly captuing the spin squeezing dynamics).  

The strategy, introduced in Refs.~\cite{zhong1993spinwave,roscilde2023rotor,roscilde2023entangling} consists in re-summing the nonlinear terms including exclusively the zero-momentum bosons to all orders. This procedure reconstructs the true nature of the $n = 0$ excitations in a finite-size system, which are not linear bosonic quasiparticles, but rather the nonlinear excitations of a macroscopic quantum rotor. The result of this procedure leads to the rotor spin-wave theory~\cite{roscilde2023rotor} 
\begin{align}
    H_{\mathcal{G}-XXZ}\approx E_{\mathrm{gs}}+\frac{K_z^2}{2\mathcal{N}_\mathcal{G}}+\sum_{n\neq 0}\omega_n\alpha_n^\dagger\alpha_n,\label{eq: H rotor-SW}
\end{align}
where $\mathbf{K} = (K_x,K_y,K_z)$ is an
angular momentum operator of macroscopic length $Ns$,
associated with the $n = 0$ bosons, namely: $K_x = Ns-a_0^\dagger a_0$, $K_y = (\sqrt{2Ns-n_0}b_0+h.c.)/2$, $K_z =(\sqrt{2Ns-n_0}b_0-h.c.)/2i$. The corresponding rotor dynamics, which describes the evolution of the zero mode, exactly reproduces the dynamics of an OAT model~\eqref{eq: OAT Hamiltonian}. The moment of inertia of the rotor variable, setting the time scale of the OAT spin squeezing, is related to the degree of the graph $\mathrm{deg}_{\mathcal{G}}$ as
\begin{align}
    \frac{1}{2\mathcal{N}_\mathcal{G}} = \mathrm{deg}_{\mathcal{G}}\frac{(1-\Delta)}{2(N-1)},
\end{align}

The spin-waves dynamics, on the other hand, occurs on the scale of the spectral gap, which only depends on the Laplacian spectral gap. Its scaling with the system size is dictated by the spectral dimension as
\begin{align}
    \delta\omega_\mathcal{G}\approx\sqrt{\delta\lambda}\approx N^{-1/d_s},
\end{align}
leading to the OAT like spin squeezing condition in Eq.~\eqref{spectral_gap_condition}(see App.~\ref{app: Rotor-Spin wave} for additional details). 

\subsection{Self averaging random graphs}
For generic random graphs, site-dependent degree fluctuations may disrupt the correspondence between the Laplacian and the adjacency matrix eigenbases. Still, for self-averaging random graphs, disorder fluctuations are suppressed int he thermodynamic limit and the hypothesis of regularity can be substituted with the one of regularity on the average. The valdity of this assumption can be tested by comparing the real space spin-wave problem, obtained by diagonalizing the $2N\times 2N$ matrix $\mathbb{H}$ in Eq.~\eqref{eq: sw Hamiltonian matrix}. Then, the real space spin waves spectrum can be compared with the one obtained by assuming self-averaging/regularity. Within the assumption of regularity on the average, the Laplacian spectrum of the regular graph is obtained from the adjacency matrix by substituting the random degree of each site with its spatial average value
\begin{align}
\mathrm{deg}_\mathcal{G}\to\overline{\mathrm{deg}_\mathcal{G}}\approx\frac{1}{N}\sum_j\deg_j = \frac{1}{N}\sum_{i,j}J_{i,j}.\label{eq: average degree}
\end{align}
As shown in Fig.~\ref{fig: SpinWave_Spectrum} this approximation correctly reproduces the spin-wave spectrum as long as the effective system size (considering the effect of  dilution) is sufficiently large. 

Thus, for self-averaging graphs, the rotor spin-wave theory captures the scaling behavior in the thermodynamic limit, resulting in the effective Hamiltonian~\eqref{eq: H rotor-SW}, where the rotor moment of inertia is now proportional to the average graph degree as
\begin{align}
    \frac{1}{2\mathcal{N}_\mathcal{G}}\approx\frac{(1-\Delta)}{2N(N-1)}\sum_{i,j}J_{i,j}\,.
\end{align}
while the spin-wave dispersion relation reads
\begin{align}
    \omega_n\approx \sqrt{\lambda_n\left(\Delta\lambda_n+\frac{1-\Delta}{N}\sum_{i,j}J_{i,j}\right)}\,.\label{eq: average degree spectrum}
\end{align}
\begin{figure}
    \centering    \includegraphics[width=0.9\linewidth]{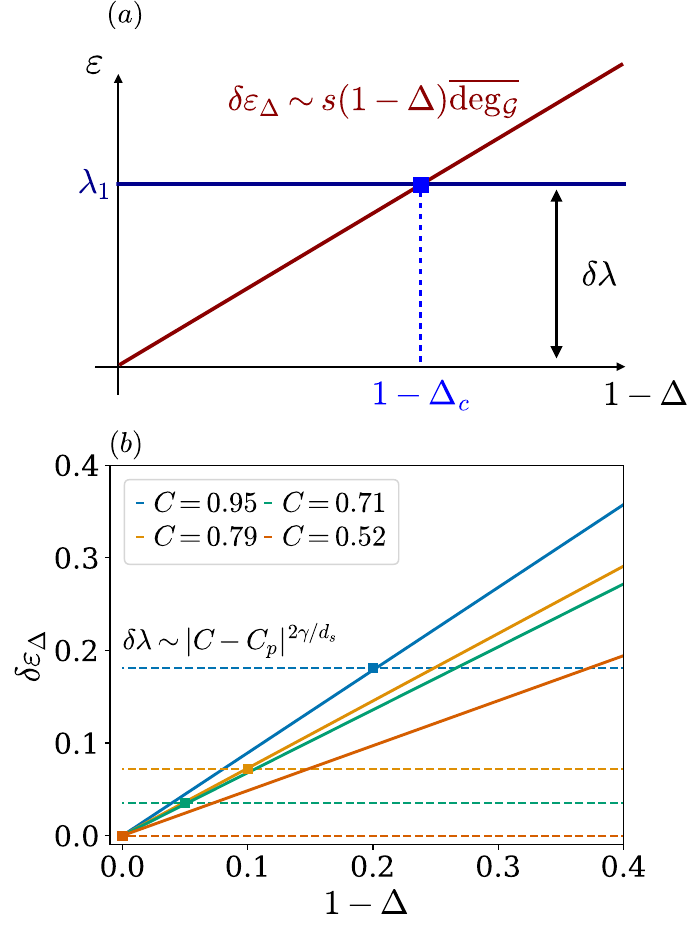}
    \caption{$(a)$ Schematic comparison of the two spectral gaps: the graph Laplacian gap $\delta\lambda$ (blue line), setting the energy scale of the unperturbed Hamiltonian at the Heisenberg point, and the anisotropy-induced gap $\delta\varepsilon_\Delta$(red line). The critical point is determined by the condition $\delta\lambda = \delta\varepsilon_{\Delta_c}$. $(b)$ Solid lines represents the perturbation energy scale $\delta\varepsilon_\Delta = s(1-\Delta)\overline{\mathrm{deg}_\mathcal{G}}$ as a function of $1-\Delta$ for different bond activation probabilities $C>C_p$ at fixed $\alpha = 1.8$, where $\overline{\mathrm{deg}_\mathcal{G}}$ is the numerically computed degree averaged over the graph nodes. Horizontal dashed lines indicate the expected scaling of the graph Laplacian spectral gap $\delta\lambda$ close to the percolation transition, as predicted by Eq.~\eqref{eq: Delta_c scaling}. Square markers denote the critical points extracted from the crossing of the minimum spin-squeezing parameter for different system sizes, using the same numerical data as in Figs.~\ref{fig: Fig_phasediagram} and~\ref{fig: Squeezing_mxy_percolation}c (see Sec.~\ref{subsec: The role of spatially correlated disorder} for additional details). The close agreement between the numerically determined critical points and the intersection of the two spectral gaps supports the perturbative criterion for $\Delta_c$.}
   \label{fig: gaps}
\end{figure}
\subsection{Perturbation theory near the Heisenberg point}\label{subsec: Perturbation theory close to the Heisenberg point at strong dilution}
Scalable spin squeezing occurs in thermodynamically large systems, so it is natural that its behavior follows the universal scaling laws dictated by rotor spin-wave theory for OAT-like squeezing and by condition C for critical squeezing. However, condition C in Sec.~\ref{sec: Conditions for scalable spin-squeezing} only establishes the universal parameters that allow crititcal squeezing of the XXZ Hamiltonian, but does not yield an estimate of the critical value $\Delta_{c}$ below which it is realized.

Since critical squeezing is a finite energy property, low-energy scaling cannot be used directly to capture the critical value of $\Delta = \Delta_c$ at a generic value of $p$. Indeed, critical squeezing follows from the existence of the finite temperature transition, which for $2<d_{s}<3$ lies in the strongly correlated regime. An estimate of the scaling of $\Delta_{c}$ with the probability of dilution $p$ or $C$ may be obtained by perturbative expansion of the spin wave Hamiltonian close to the Heisenberg point ($\Delta = 1$). The real space Hamiltonian in Eq.~\eqref{eq: sw Hamiltonian matrix} can be written as
\begin{align}
    \mathbb{H} = \mathbb{H}_{\Delta = 1}+\mathbb{V},
\end{align}
where $\mathbb{H}_{\Delta = 1}$ is the unperturbed Hamiltonian at $\Delta = 1$ 
\begin{align}
\mathbb{H}_{\mathrm{\Delta = 1}} = \begin{pmatrix}
    s\mathbb{L} &\mathbb{O}\\
    \mathbb{O} &s\mathbb{L}
    \end{pmatrix}\label{eq: sw Hamiltonian matrix Heisenberg}
\end{align}
and $\mathbb{L}=\mathbb{D}-\mathbb{J}$ is the graph Laplacian~\eqref{eq: graph Laplacian}.
This reflects the fact that the Heisenberg Hamiltonian acts as the Laplacian in the single-excitation subspace spanned by the states obtained by flipping a single spin from the ground state  $\ket{\phi_m} = \ket{00\dots01_m0\dots0}$.
The spectrum of $\mathbb{H}_\Delta$ corresponds to the Laplacian spectrum $\{\lambda_n\}$ and it's degenerate with eigenvectors of the form
\begin{align}
    \begin{pmatrix}
        \mathbf{u}_n\cos\theta\\
        \mathbf{u}_n e^{\varphi}\sin\theta
    \end{pmatrix},\label{eq: degenerate eigenspace}
\end{align}
reflecting the $SU(2)$ symmetry of the Heisenberg point. The degeneracy is lifted by the perturbation 
\begin{align}
    \mathbb{V}_{\Delta}=\frac{s(1-\Delta)}{2}
    \begin{pmatrix}
        \mathbb{J} &-\mathbb{J}\\
        -\mathbb{J} &\mathbb{J}
    \end{pmatrix},
\end{align}
which has two eigenvalues $\{s(1-\Delta)\mathbf{u}_n^{T}\cdot\mathbb{J}\cdot\mathbf{u}_n, 0\}$ within each degenerate subspace. Accordingly, we obtain the energy levels splitting
\begin{align}
    \omega_{n,0} = \lambda_n,\quad\omega_{n,1} = \lambda_n+\varepsilon_{n,\Delta},
\end{align}
where, applying degenerate perturbation theory and keeping corrections up to leading order in $(1-\Delta)$, we have that
\begin{align}
\varepsilon_{n,\Delta} 
&=s(1-\Delta)\mathbf{u}_n^T\cdot\mathbb{J}\cdot\mathbf{u}_n\notag.%\\
%&+s^2(1-\Delta)^2\sum_{m\neq n}\frac{|\mathbf{u}_m^T\cdot\mathbb{J}\cdot\mathbf{u}_n|^2}{\lambda_n-\lambda_m}+\mathcal{O}((1-\Delta)^3)
\end{align}
In particular the leading correction to the ground state reads
\begin{align}
   \varepsilon_{0,\Delta} = s(1-\Delta)\frac{1}{N}\sum_{i,j}J_{i,j} \approx s(1-\Delta)\overline{\mathrm{deg}_\mathcal{G}},\label{eq: epsilon_0_Delta}
\end{align}
where $\overline{\mathrm{deg}_\mathcal{G}}$ is the average degree of the graph~\eqref{eq: average degree}. Therefore, as shown in Fig.~\ref{fig: gaps} two energy scales emerge corresponding to the spectral gaps
\begin{align}
    \delta\lambda &= \omega_{1,0}-\omega_{0,0}\,,\\
    \delta\varepsilon_\Delta &= \omega_{0,1}-\omega_{0,0} \approx s(1-\Delta)\overline{\mathrm{deg}_\mathcal{G}}\,.
\end{align}
Comparing these two energy scales we obtain the condition 
\begin{align}
    \delta\lambda \approx s(1-\Delta)\overline{\mathrm{deg}_\mathcal{G}}\,.\label{eq: dlamb-1-Delta}
\end{align}
Given the fact that at the percolation threshold $p\to p_p$ the average degree is $\overline{\mathrm{deg}_\mathcal{G}}\sim \mathcal{O}(1)$, Eq.~\eqref{eq: dlamb-1-Delta} leads to the condition~\eqref{eq: delta_eps-delta_lamb}
used in Sec.~\ref{subsec: Scaling of the critical point in proximity to the percolation threshold} to estimate the critical point scaling in the proximity of the percolation threshold. 

\section{Application to experiments}\label{sec: Application to experiments}
In this section, we connect the theoretical framework developed above to experimentally relevant platforms, focusing on the realization of scalable spin squeezing in long-range interacting spin models on different network geometries and affected by different types of disorder. The setups we consider, the corresponding experimental platforms and the results for the conditions on scalable spin squeezing identified by specializing the general results of the previous section are summarized in Tab.~\ref{tab: Summary application to experiments}.

\subsection{Spatially uncorrelated disorder}\label{subsec:Spatially uncorrelated disorder}
\begin{figure*}
    \centering    \includegraphics[width=\linewidth]{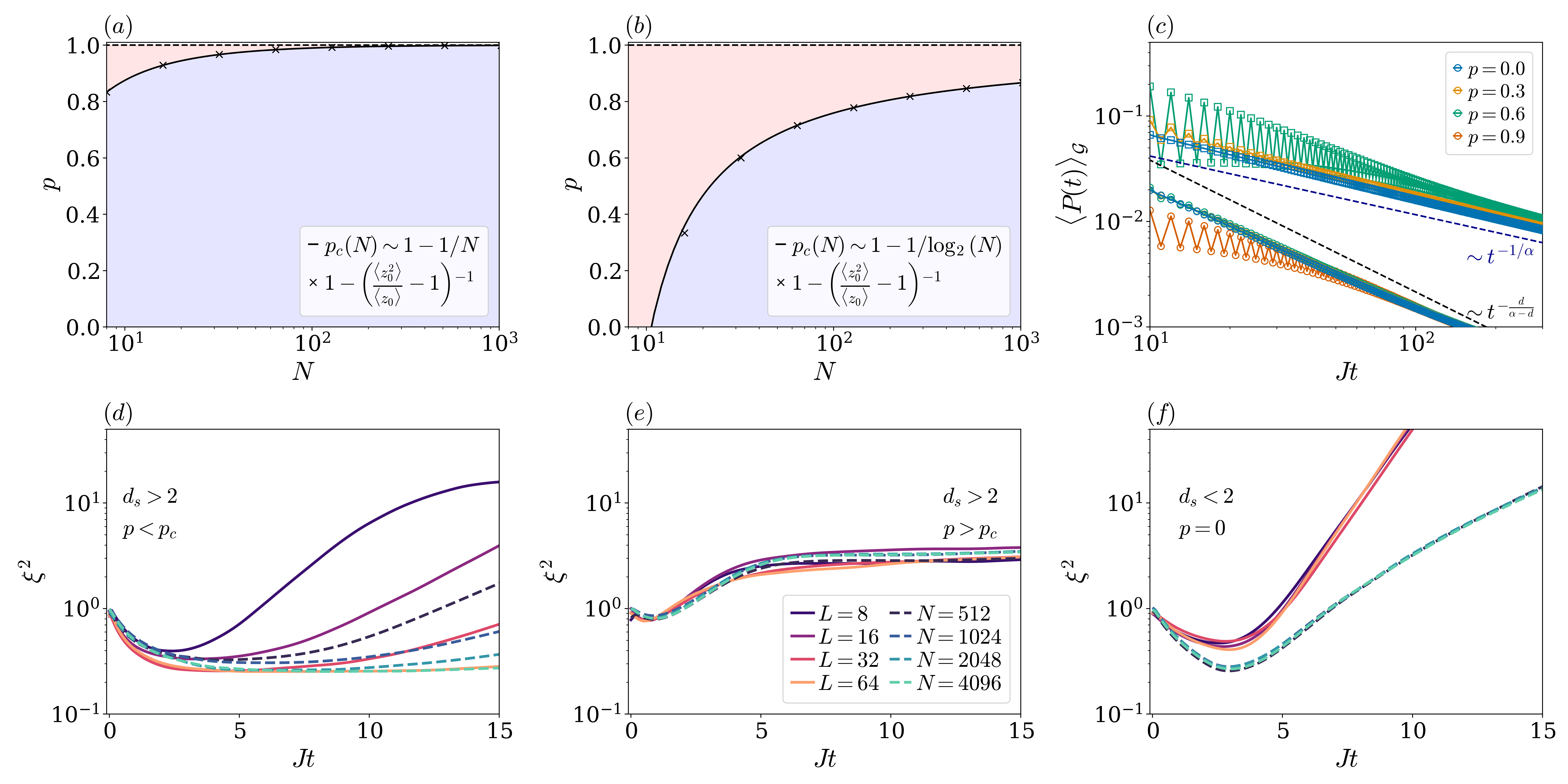}
    \caption{\textbf{Results for spatially uncorrelated disorder}: $(a)$-$(b)$ Finite-size percolation phase diagram for a long-range diluted lattice (a) and a power-of-two (PW2) graph $(b)$. Black crosses denote the numerically determined percolation threshold obtained from Eq.~\eqref{eq: critical p}, while the black dashed line shows the analytical prediction in the in the large $N$ limit~\eqref{eq: critical p LR diluted} and~\eqref{eq: p_p PW2}. $(c)$  Random walk recurrence probability averaged over the nodes $\langle P(t)\rangle_\mathcal{G}$ as function of time in a long-range diluted chain (dots) and a PW2 graph (squares) for $\alpha = 1.8$, and for different site dilution probabilities $p = 0,0.3,0.6,0.9$, the system size is $N = 2500$. The numerical data are compared to clean system system prediction~\eqref{eq: recurrence probability ds} with spectral dimensions~\eqref{eq: spectral dimension LR diluted} (black dashed line) and \eqref{eq: ds PW2} (blue dashed line). (d)-(f) Time evolution of the spin-squeezing parameter $\xi^2$ for different system sizes (darker shades correspond to larger $N$), for different geometries, interaction exponent $\alpha$, anisotropy $\Delta$ and dilution probability $p$, obtained by DTWA simulations (see App.~\ref{app: DTWA}). Solid lines show the numerical results for a two-dimensional triangular lattice with $\alpha = 3$ (spectral dimension $d_s = 4>2$); $\Delta = 0$ and $p = 0.2$ (panel $(d)$), $\alpha = 3.8$ ($2<d_s\simeq 2.22<3$), $\Delta = -0.8$ and $p = 0.5$ (panel $(e)$); a one-dimensional lattice with $\alpha = 3$ ($d_s = 1<2$) in the absence of disorder $p=0$ (panel $(f)$). Dashed lines show the numerical results for a PW2 graph with $\alpha = 0.5$ (spectral dimension $d_s = 4>2$), $\Delta = 0$ and $p = 0.2$ (panel $(d)$); $(e)$ $\alpha = 0.9$ ($2<d_s\simeq 2.22<3$), $\Delta = -0.8$ and $p = 0.5$ (panel $(e)$); $\alpha = 1.2$ ($d_s = 5/3<2$) in the absence of disorder $p=0$ (panel $(f)$).}\label{fig: Fig_examples}
\end{figure*}
Let us start from the case of spatially uncorrelated disorder characterised by a bond probability $p_{i,j} = p_ip_j$. In the experimental scenario we consider here disorder is typically modeled as uncorrelated site dilution, where each site is removed independently with probability $p$ (see Fig.~\ref{fig: Fig1_schematic}a-b).

In this case the emergence of a percolating cluster is controlled by the first two moments of the distribution of the number of connected neighbors (also referred to as degree distribution in the graph theory terminology) of the undiluted graph, $\langle z_0\rangle$ and $\langle z_0^2\rangle$~\cite{cohen2000resilience,callaway2000robustness,albert2000error}. As shown in Ref.~\cite{cohen2000resilience}, the critical percolation threshold can be estimated as
\begin{align}
    p<p_p \approx 1-\left(\frac{\langle z_0^2\rangle}{\langle z_0\rangle}-1\right)^{-1}.\label{eq: critical p}
\end{align}
For $p<p_p$ a connected cluster of size scaling with $N$ exists, and the scalable spin squeezing remains possible despite a finite filling fraction, although with a reduced effective system size of $\sim (1-p)N$. If instead $p>p_p$, the system decomposes into disconnected clusters of order one, and the long-range correlations required for squeezing cannot develop. In clean lattice or graph geometries with long-range interactions, the nodes have all-to-all connectivity, and the moments of the degree distribution grow with $N$ in the thermodynamic limit, implying
\begin{align}
\lim_{N\to\infty}p_p\to 1,\end{align}
and guaranteeing percolation in infinite long-range interacting site diluted graphs for any finite dilution $p<1$. 
However, for any finite $N$ a critical dilution probability above which no giant percolating cluster exists, preventing the onset of scalable spin squeezing. As shown below, the finite size corrections strongly depend on the underling clean graph structure. Figures~\ref{fig: Fig_examples}a and~\ref{fig: Fig_examples}b show the size percolation phase diagrams for the two examples considered below of a site diluted long-range lattice and a site diluted power-of-two graph, respectively. This celarly shows that the more sparse power-of-two graph structure is more prone to finite size effects.   

As shown in Sec.~\ref{sec: Conditions for scalable spin-squeezing}, from the graph-theoretic perspective, the key quantity controlling scalable squeezing is the spectral dimension $d_s$. A central result of our analysis is that $d_s$ is unaffected by site dilution: long-wavelength collective modes, which dominate squeezing dynamics, remain insensitive to local defects as long as a percolating cluster and an ordered phase exist.
This can be seen from the long time behavior of the recurrence probability~\eqref{eq: recurrence probability ds}. As shown in Fig.~\ref{fig: Fig_examples}a even at very high dilution probabilities $p = 0.9$, as long as the system size is sufficiently large, the long time behavior of the recurrence probability converges to~\eqref{eq: recurrence probability ds} with the same spectral dimension as for the disorderless system.

Thus, contrary to early expectations~\cite{block2024scalable}, a constant dilution does not affect the possibility of achieving scalable OAT-like squeezing ($d_s>3$), at least as long as the percolation condition remains satisfied, which is always the case at large-$N$. Moreover, for critical squeezing ($2<d_s<3$), we also need the anisotropy $\Delta$ is such that the initial state energy is sufficiently low to allow for $xy$-ferromagnetic order. Let us now analyze more in detail the two examples of site diluted geometries. 

\subsubsection{Diluted long-range lattices}\label{subsec: Diluted long-range lattices}
We start by considering systems in which spin degrees of freedom occupy the nodes of a regular lattice and interact via long-range couplings decaying as a power-law of the intersite distance $r_{ij}$: $J_{i,j} \propto {r_{ij}}^{-\alpha}$. Spatial disorder is introduced through random site dilution, where each site is occupied with equal probability $f = 1-p$. Physically, this corresponds to situations in which a fraction of the atoms, ions, or molecules that form the effective two-level system are lost or decoupled from the coherent dynamics. Such imperfect filling occurs naturally in a broad range of quantum simulators, including trapped-ions~\cite{britton2012engineered, kiesenhofer2023controlling, guo2024siteresolved}, Rydberg atoms~\cite{browaeys2020manybody,gross2017quantum}, ultracold molecules~\cite{moses2015creation,holland2023ondemand}, and NV-centers~\cite{doherty2013nitrogen,gong2023coherent,hughes2025strongly}.

At sufficiently small $\alpha$, the disorder-free model exhibits scalable spin squeezing when the $\Delta$ anisotropy lies within the $xy$-ferromagnetic phase. The region of $\alpha$ where scalable squeezing occurs is readly obtained from the the spectral dimension $d_s$, which in long-range lattices is related to the interaction exponent $\alpha$ by~\cite{kotliar1983onedimensional,banos2012correspondence,angelini2014relations,behan2017scaling, defenu2017criticality,solfanelli2024universality,solfanelli2025universalwork}
\begin{align}
d_s = \frac{2d}{\alpha - d},\label{eq: spectral dimension LR diluted}
\end{align}
with $d$ the physical spatial dimension of the lattice (see App.~\ref{app: Spectral dimension of long-range lattices}). OAT like squezing is then found for $\alpha<5d/3$ according to the discussion below Eq.~\eqref{duc_condition}. 

The critical regime, where squeezing occurs due to the existence of the finite temperature transition and with scaling exponents different from the OAT ones, extends beyond $\alpha=5d/3$. According to condition B in Sec.~\ref{sec: Conditions for scalable spin-squeezing}, the onset of critical squeezing and spontaneous symmetry breaking share the same threshold, i.e., $\alpha<2d$, as obtained from Eq.~\eqref{critical_squeezing_window} and consistent with the findings of Ref.~\cite{block2024scalable} for clean lattices. Moreover, as shown in App.~\ref{app: Role of an exponential cutoff}, these results are robust against the presence of an exponential cutoff on the tails of the interaction profile $J_r\propto r^{-\alpha}e^{-\kappa r}$, as long as long as the decay rate $\kappa$ scales with the system size as $\kappa\sim 1/N$. This is a typical scenario in trapped ions experiments~\cite{monroe2021programmable,schuckert2025observation}.

This insight is readily generalized to diluted long-range lattices, since the percolation thresholds saturates in the thermodynamic limit ($\lim_{N\to\infty}p_c=1$) and therefore the existence of a giant cluster is guaranteed for any $\alpha$ and $\Delta$. Still, uncorrelated disorder shifts the effective percolation threshold at finite $N$
\begin{align}
p_p(N) = 1-\frac{1}{N-2},\label{eq: critical p LR diluted}
\end{align}
as predicted by the general result in Eq.~\eqref{eq: critical p}. This leads to finite-size corrections of order $\mathcal{O}(N^{-1})$ that suppress squeezing in smaller systems. The resulting finite-size percolation phase diagram is illustrated in Fig.~\ref{fig: Fig_examples}a. Intuitively, the uncorrelated disorder effectively reduces the number of sites available to squeezing $N_p = (1-p)N$. Then, for any $N>2$, the critical percolation probability~\eqref{eq: critical p LR diluted} leads to an effective system size $N_{p_c}<2$. Therefore, in absence of the giant cluster, the system is effectively built by isolated independent spins, which cannot develop quantum correlations and, therefore, cannot form a macroscopic squeezed state.

For giant clusters close to the critical threshold $p\lesssim p_p(N)$, we can apply Eq.~\eqref{eq: Delta_c scaling} to find out how the critical value of the anisotropy $\Delta_c$ scales close to the Heisenberg point. Since  site diluted lattices have all-to-all connectivity, the percolation transition belongs to the mean field universality class with $\gamma = 1$. Moreover, the spectral dimension in Eq.~\eqref{eq: spectral dimension LR diluted} is not affected by site dilution (as long as we are within the percolating phase), this can be seen from the long time behavior of the recurrence probability shown in Fig.~\ref{fig: Fig_examples}c. Accordingly, we find the scaling
\begin{align}
    (1-\Delta_c)\sim |p_p(N)-p|^{\alpha/d-1}\,.\label{eq: Deltac p LR diluted}
\end{align}

The validity of the universal conditions for squeezing is readily demonstrated by a numerical study of the experimentally relevant example of a two-dimensional triangular lattice (see Fig.~\ref{fig: Fig1_schematic}a) with power-law interactions $J_{i,j}=r_{i,j}^{-\alpha}$ and finite site dilution probability $p$. We consider a power law decay exponent $\alpha = 3$, typical of any dipolar interacting system in $d =2$ including, ultracold molecules~\cite{moses2015creation,holland2023ondemand} and solid state systems~\cite{doherty2013nitrogen,gong2023coherent,hughes2025strongly}, where limited filling fraction typically plays a particularly relevant role~\cite{wu2025spinsqueezing}. 

The solid lines in Figs.~\ref{fig: Fig_examples}d-f, compare the time evolution of the spin squeezing parameter $\xi^2$~\eqref{eq: squeezing parameter}
for different system sizes $N$ and dilution probabilities $p$ and anisotropy $\Delta$, obtained using the DTWA method (see App.~\ref{app: DTWA}). As expected from our theory, spectral dimension $d_s>2$ and anisotropy sufficiently close to the Heisenberg point ($\Delta>\Delta_c(p,\alpha)$, as estimated from Eq.~\eqref{eq: Deltac p LR diluted}), we observe a 
minimum of $\xi^2$ that decreases with the system size scalable squeezing (Fig.~\ref{fig: Fig_examples}d). On the other hand, for $2<d_s<3$ and  $\Delta<\Delta_c(p,\alpha)$, long-time order is destroyed and the minimum of $\xi^2$ becomes size independent, indicating the absence of scalable squeezing (Fig.~\ref{fig: Fig_examples}e). Finally, for $d_s<2$, the absence of scalable squeezing and the destruction of $xy$-order signal the impossibility of scalable metrological advantage even at zero dilution probability $p=0$ (see Fig.~\ref{fig: Fig_examples}f).

\subsubsection{Graph geometries}\label{subsec: Hypergraphs structures}
Going beyond diluted lattice geometries, we now consider how spin-squeezing dynamics is affected by more complex interaction graphs whose dimensionality, topology, and metric structure are entirely distinct from the physical embedding of the atomic array.

Such graph geometries can be experimentally realized in setups of cold atoms trapped in optical tweezers inside an optical cavity and subject to spatially varying magnetic fields~\cite{periwal2021programmable}. In particular, programmable spin-exchange interactions can be engineered by placing an array of atoms positioned in optical tweezers and excited to Rydberg states within a single-mode optical cavity.
While the cavity mode naturally mediates all-to-all interactions, this connectivity can be selectively broken by introducing a magnetic field gradient along the cavity axis. The gradient induces an energy mismatch $\hbar \omega_B$ 
between the Zeeman splittings of adjacent sites, rendering spin-exchange processes between distant ensembles off-resonant. Interactions between ensembles separated by a distance of $r$ sites can then be selectively restored by modulating the intensity of the drive field at frequency $r\omega_B$, effectively engineering a time-dependent spin-exchange coupling with programmable spatial structure~\cite{periwal2021programmable}.

An experimentally relevant example of such engineered graph geometries is the so-called \emph{power-of-two} (PW2) graph, in which the coupling matrix elements are defined as~\cite{periwal2021programmable,bentsen2019treelike}
\begin{align}
    J_{i,j} =
    \begin{cases}
        |i-j|^{-\alpha}, & \text{for } |i-j| = 2^{n},\\
        0, & \text{otherwise}.
    \end{cases}
\end{align}
This geometry has attracted significant attention in the context of quantum information spreading~\cite{hashizume2021deterministic,hashizume2022tunable,hashizume2022measurement}. In particular, for $\alpha = 0$ it gives rise to fast scrambling dynamics and black-hole-like behavior~\cite{bentsen2019treelike}. More recently, it has been shown that in such mean-field limit $\alpha=0$ the PW2 graph supports one-axis-twisting-like scalable spin squeezing, associated with a gapped spin-wave spectrum in the thermodynamic limit and a diverging spectral dimension $d_s\to\infty$~\cite{kuriyattil2025entangled} (see App.~\ref{app: Expansion of the power-of-two graph spectrum}). 
Here we instead focus on the physically distinct regime $\alpha\neq 0$, where the graph hosts a nontrivial and finite spectral dimension $d_s<\infty$.

In the absence of disorder, the power-of-two graph is translationally invariant, $J_{i,j}=J(|i-j|)$, allowing its spectrum to be analyzed in Fourier space. The Fourier transform of the coupling matrix, which determines the single-particle dispersion of spin-wave excitations, reads
\begin{align}
    \tilde{J}_k = \sum_{n=0}^{\log_2(N)-1} 2^{-n\alpha}\cos(2^{n}k),\label{eq: PW2 Jk}
\end{align}
where $k$ labels the quasiparticle momentum modes.

In the $N\to\infty$ limit, and for $0<\alpha<2$, $\tilde{J}_k$ corresponds to the so called Weierstrass function~\cite{weierstrass1988,hardy1916weierstrass,hunt1998thehausdorff,david2018bypassing}. This function is continuous everywhere but nowhere differentiable, and exhibits a self-similar fractal structure reflecting the recursive properties of the interaction graph. As discussed in App.~\ref{app: Expansion of the power-of-two graph spectrum}, despite the absence of a well-defined derivative, it is still possible to meaningfully define a spectral dimension governing the universality of spin systems in the PW2 geometry. 

For $\alpha>0$ this is achieved expanding the spectrum near the zero mode $k\simeq 0$, corresponding to the uniform Laplacian ground state $\lambda_0 = 0$, as the second Laplacian eigenvalue $\lambda_1$ corresponds to the lowest momentum $\lambda_1= k_1^2= (2\pi/N)^2$. The resulting low-energy dispersion relation agrees with the rigorous bounds of the Weierstrass function~\cite{david2018bypassing} and takes the form
\begin{align}
    \tilde{J}_0-\tilde{J}_k\approx\begin{cases}
        k^2 &\mathrm{if}\,\,\alpha>2\\
        k^{\alpha}&\mathrm{if}\,\,0<\alpha<2
    \end{cases}\,.\label{eq: PW2 dispersion}
\end{align} 
This directly leads to the spectral dimension 
\begin{align}
    d_{s,\mathrm{PW}2} = \begin{cases}
        1 &\mathrm{if}\,\,\alpha\geq2\\
        2/\alpha&\mathrm{if}\,\,0<\alpha<2
    \end{cases}\label{eq: ds PW2 positive alpha}
\end{align}

For interactions growing with the distance $\alpha<0$, a normalization factor has to be introduced in the couplings $J_{i,j}\to N^{\alpha}J_{i,j}$ in order to ensure energy extensivity and a meaningful thermodynamic limit~\cite{kac1963van}. The strongest bonds for $\alpha<0$ are those at the largest distances ($r \sim N/2$), effectively inverting the hierarchy of energy scales. The ground state remains the uniform mode ($\lambda_0=0$). However, the first excited state is no longer the long-wavelength mode $\lambda_1=k_1^2$, but the staggered mode $\lambda_{N/2} = k_{N/2}^2 = \pi^2$. This staggered mode satisfies the strong longest-range bonds (which contribute positively to $J_{\pi}$ for large $n$) and is only frustrated by the weakest short-range bonds. The spectral gap is dominated by the scaling of the normalization factor $ N^\alpha$ leading to
\begin{equation}
    \delta\lambda\approx 2 N^{\alpha}\left( 1 - \cos(\pi) \right) \sim N^{\alpha} = N^{-|\alpha|}.
\end{equation}
This allows to extend Eq.\eqref{eq: ds PW2 positive alpha} to generic values of $\alpha$ obtaining the spectral dimension
\begin{align}
    d_{s,\mathrm{PW}2} = \begin{cases}
        1 &\mathrm{if}\,\,\alpha\geq2\\
        2/|\alpha|&\mathrm{if}\,\,\alpha<2
    \end{cases}\label{eq: ds PW2}\,.
\end{align}
 Since condition B in Sec.~\ref{sec: Conditions for scalable spin-squeezing} translates to $d_{s,\mathrm{PW2}}>2$ for the PW2 graph, one finds that scalable squeezing occurs for the PW2 graph when $|\alpha|<1$ according to Eq.~\eqref{eq: ds PW2}.

As for regular lattices, the effect of a finite dilution probability on the spectral dimension is encoded in the long-time behavior of the random walk recurrence probability. For sufficiently low dilution probability after some transient oscillations, this long time behavior converges to the clean system prediction~\eqref{eq: recurrence probability ds} (blue dashed line) signalling the irrelevance of uncorrelated site disorder (see square points in Fig.~\ref{fig: Fig_examples}c).

On the other hand, differently from diluted long-range lattices for high dilution probability $p = 0.9$ (not shown), $\langle P(t)\rangle_\mathcal{G}$ keeps oscillating and does not converge to the expected power law behavior, within the time window defined by the spectral gap time scale $t<t^*= N^{2/d_s}$. Indeed, differently from the case of diluted regular lattices, the degree moments of the PW2 graph only diverge logarithmically with system size. Due to this fact, finite size fluctuations are exponentially stronger in the PW2 graph with respect to the lattice case. Accordingly, the percolation threshold saturates logarithmically  in the thermodynamic limit
\begin{align}
    p_c(N) = 1 - \frac{1}{\log_2(N)-5/2-1/N}.\label{eq: p_p PW2}
\end{align}
For the PW2 graph the limit $p_c(N)\to 1$ is only reached  later as $N\to\infty$ with respect to the case of diluted regular lattices, as it emerges from the comparison between Fig.~\ref{fig: Fig_examples}a and Fig.~\ref{fig: Fig_examples}b. 

Characterizing the scaling of $1-\Delta_c$ in proximity of the percolation threshold for PW2 is done in analogy to the case of the regular lattice. Inserting in Eq.~\eqref{eq: Delta_c scaling} the explicit expression for the spectral dimension of the PW2 graph~\eqref{eq: ds PW2} and accounting for the mean-field scaling of the percolation problem ($\gamma =1$) leads to
\begin{align}
    1-\Delta_c\sim |p_p(N)-p|^{\alpha}\,.\label{eq: 1-Deltac PW2}
\end{align}
With respect to the regular lattice case, the scaling of the critical anisotropy in the vicinity of the percolation threshold for the PW2 graph approaches the Heisenberg point faster (with an exponent $\alpha$ instead od $\alpha-1$ in $d=1$), resulting in larger impact of dilution on the realization of scalable spin squeezing. 

Figures~\ref{fig: Fig_examples}d-f, compare the time evolution of the spin squeezing parameter $\xi^2$~\eqref{eq: squeezing parameter} in the PW2 graph geometry (dashed lines) with respect to the one of a long-range diluted lattice (solid lines), for varying system size $N$, interaction exponent $\alpha$, anisotropy $\Delta$, and dilution probability $p$, obtained via the DTWA method (see App.~\ref{app: DTWA}). The behavior closely parallels that observed for long-range diluted lattices (solid lines in Figs.~\ref{fig: Fig_examples}d–f), once the data are compared at equal values of the spectral dimension $d_s$~\eqref{eq: ds PW2}. In particular, for $d_s>2$ ($|\alpha|<1$) and $\Delta>\Delta_c(p,\alpha)$ we observe scalable spin squeezing (Fig.~\ref{fig: Fig_examples}d). For $d_s>2$ but $\Delta<\Delta_c(p,\alpha)$, no scalable spin squeezing is achieved (Fig.~\ref{fig: Fig_examples}e). Finally, when $d_s<2$ ($|\alpha|>1$), spontaneous symmetry breaking is precluded on the PW2 graph, and scalable spin squeezing cannot be realized even in the absence of spatial disorder (Fig.~\ref{fig: Fig_examples}f). 
\begin{figure}
    \centering    \includegraphics[width=0.9\linewidth]{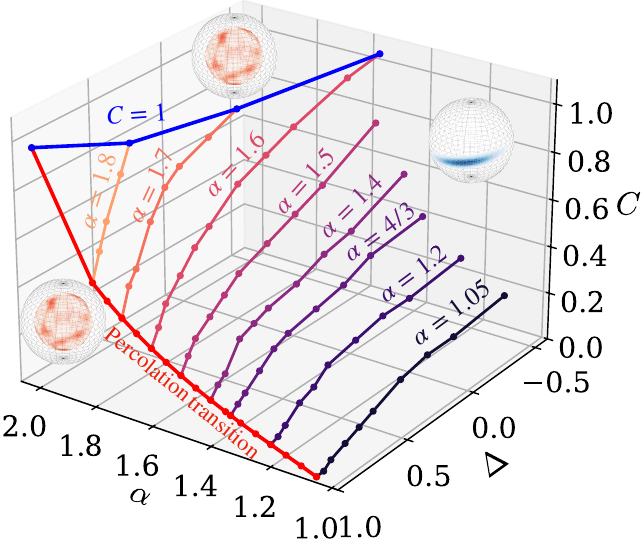}
    \caption{Spin-squeezing phase diagram as a function of the long-range exponent $\alpha$, anisotropy $\Delta$ and the bond probability $C$, for spatially correlated disorder with bond dilution probability given by Eq.~\eqref{eq: pij LR ring bis}. The critical disorder strength $C_c(\alpha,\Delta)$ depends explicitly on $\alpha$ and on the anisotropy $\Delta$ reflecting the disorder-induced shift of the xy-ferromagnetic critical point. For $1<\alpha<2$, a finite percolation threshold $C_p(\alpha)$ exists. Precise Monte Carlo estimates of the percolation critical point as a function of $\alpha$, from Ref.~\cite{defenu2017percolation}, are shown as the red curve in the $\Delta = 1$ plane. No scalable squeezing occurs for $C<C_p$, and as $C\to C_p$, squeezing is restricted to values of $\Delta$ increasingly close to the Heisenberg point $\Delta=1$. For $C=1$, the phase diagram is consistent the the disorder-free case~\cite{block2024scalable}. MPS data from~\cite{block2024replicationdata}, for the disorder-free case, are shown as the blue curve in the $C=1$ plane. In this regime, no scalable spin squeezing is possible for $\Delta<\Delta_c$. Numerical data are obtained from DTWA simulations (see App.~\ref{app: DTWA}). The transition $C_c(\alpha,\Delta)$ is extracted by comparing system sizes, $N=256,512,1024$, and identifying the threshold below which the (sufficiently long-time) minimum of the squeezing parameter~\eqref{eq: squeezing parameter} becomes size independent. Data are averaged over $500$ disorder and DTWA realizations; colors correspond to different $1<\alpha<2$.}\label{fig: Fig_phasediagram}
\end{figure}
\subsection{The role of spatially correlated disorder}\label{subsec: The role of spatially correlated disorder}
\begin{figure*}
    \centering    \includegraphics[width=\linewidth]{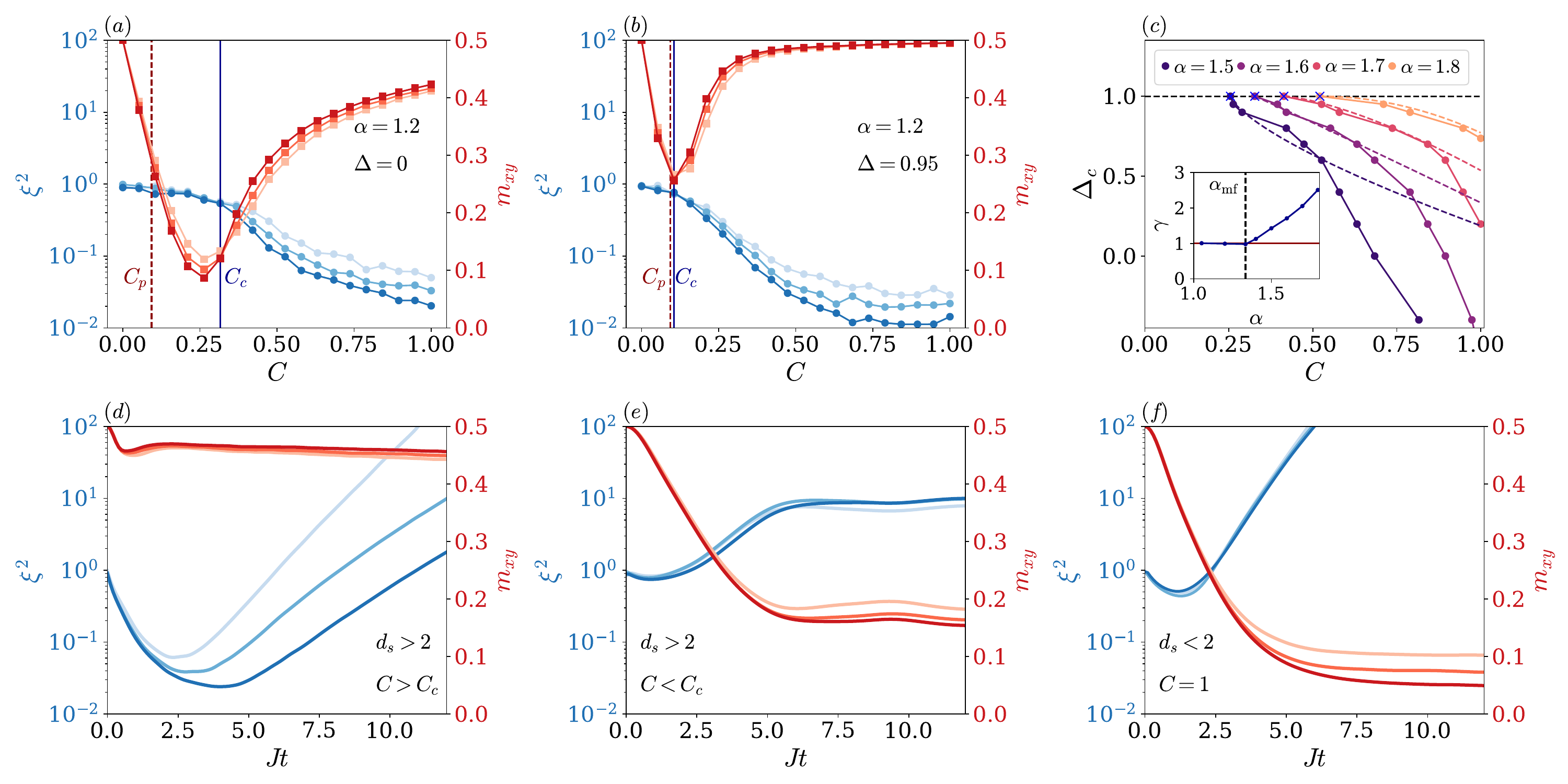}
    \caption{\textbf{Numerical study of squeezing on a lattice with power-law correlated bond probability}: $(a)$-$(b)$ Optimal spin squeezing parameter $\xi^2$ (left blue axis) and long time $xy$-magnetization $m_{xy}$ as a function of the bond activation probability $C$ for different system sizes $N=256,512,1024$ and different values of the anisotropy $\Delta = 0$ $(a)$ and $\Delta = 0.95$ $(b)$ in a lattice with long-range correlated disorder with the bond probability in Eq.~\eqref{eq: pij LR ring bis} and power law exponent $\alpha = 1.2$. As  $\Delta$ increases towards the Heisenberg point ($\Delta \to 1$) the critical probability $C_c$, below which scalable spin squeezing is attained approaches the percolation threshold $C_p$. $(c)$ Critical anisotropy as a function  of $C$ for different values of $\alpha$ corresponding to different colors. The dots are the same DTWA numerical data as for Fig.~\ref{fig: Fig_phasediagram}, the blue crosses represent the percolation threshold $C_p$ the dashed lines represent the theoretical prediction for the $C\to C_p$ and $\Delta\to 1$ behavior in Eq.~\eqref{eq: Delta_c C}. The inset shows the critical exponent $\gamma$ of long-range percolation as a function $\alpha$, for $\alpha<\alpha_{\mathrm{mf}}=4/3$ it takes the mean field value $\gamma_{\mathrm{mf}} = 1$, while beyond the mean field threshold it shows a non-trivial alpha dependent behavior. $(d)$-$(f)$ Time evolution of the spin-squeezing parameter $\xi^2$ (blue left axis) and $xy$-magnetization per spin $m_{xy}$ (red right axis) for different system sizes (darker shades correspond to larger $N = 512, 1024, 2048, 4096$), for different values of $\alpha$, $\Delta$ and $C$, obtained by DTWA simulations (see App.~\ref{app: DTWA}). $(d)$ $\alpha = 1.2$ ( $d_s = 10>2$), $\Delta = 0$ and $C\simeq 0.8 $: scalable spin squeezing and a finite long-time magnetization $m_{xy}$, indicate that $C>C_c$. $(e)$ Same geometry and parameters as in $(d)$ but with $C\simeq 0.16$: the absence of scalable squeezing and the vanishing of the long-time $m_{xy}$, in the thermodynamic limit signal are consequences of the absence of a percolating cluster $C<C_c$. $(f)$ $\alpha = 2.2$ ($d_s = 5/3<2$) and $C=1$: no scalable spin squeezing is observed, consistent with the impossibility of spontaneous symmetry breaking for $d_s<2$. }
   \label{fig: Squeezing_mxy_percolation}
\end{figure*}
In the previous examples, we demonstrated that the impact of uncorrelated disorder on the realization of scalable spin squeezing strongly depends on the underlying geometrical structure. Nevertheless, the threshold on the decay of the interaction strength $\alpha$, which generates scalable squeezing, remains unaffected by uncorrelated disorder in the thermodynamic limit.

In this section we consider a random graph structure modeling the effect of spatially correlated disorder such that the bond probability between site $i$ and $j$ cannot be written as the product of two independent probabilities for each site, i.e., $p_{i,j}\neq p_ip_j$. This may occur in any experimental setup due to the presence of spurious interactions and correlations among the sources of errors or due to interactions among the graph nodes mediated by an external environment \cite{aharonov2006faulttolerant}. This is particularly relevant to digital quantum simulators with limited connectivity, such as superconductive qubits platforms. There, connecting qubits at large distances implies a larger depth of the quantum circuit, resulting in larger error for longer bonds~\cite{solfanelli2024stabilization,weaving2023benchmarking}.

As a prototypical example of spatially correlated disorder we introduce a random graph corresponding to a fully-connected network whose bonds are erased with probability
\begin{align}
    p_{i,j} = 1-\frac{C}{|i-j|^\alpha},\label{eq: pij LR ring bis}
\end{align}
which can be interpreted as the probability of a gate failure for a two qubits gate connecting the qubits sitting at site $i$ and $j$.

The analysis of the percolation problem on this graph~\cite{defenu2017percolation} reveals that the universality class of the percolation transition is affected by the decay exponent $\alpha$.  In particular, a finite critical percolation probability $C_c(\alpha)$ exists in the thermodynamics limit:
\begin{align}
   C>C_c(\alpha)\geq 1-\frac{1}{2\zeta(\alpha)},
\end{align}
where the last inequality is a bound obtained using the exact solution on the Bethe-lattice~\cite{schulman1983longrange}.
The $\alpha$-dependent percolation threshold yields a disorder dependent phase diagram for scalable spin squeezing. Then, scalable spin squeezing is obtained only in a complex region of the $C,\alpha,\Delta$-phase space; see Fig.~\ref{fig: Fig_phasediagram}. In particular, scalable spin squeezing can be achieved for values of $C$ above the percolation threshold corresponding to the red curve in Fig.~\ref{fig: Fig_phasediagram}.

The non mean-field nature of the percolation transition in the case of correlated disorder, does not only lead to decay dependent threshold $C_c(\alpha)$, but also to a non-trivial critical scaling. Indeed, in proximity of the percolation transition $C_p\lesssim C$, the general result in Eq.~\eqref{eq: Delta_c scaling} leads to the scaling 
\begin{align}
    (1-\Delta_c)\sim [C-C_p(\alpha)]^{\gamma(\alpha/d-1)},\label{eq: Delta_c C}
\end{align}
which, differently from the previous cases, features a disorder dependent power-law exponent $\gamma$. The latter is tied to the universality class of the percolation problem and to the value of the spectral dimension of the model~\cite{millan2021complex,bighin2024universal}. In particular, for $\alpha\leq 4/3$, the spectral dimension is above the upper critical dimension for percolation $d_s\geq 6$. There, the percolation transition is in the mean-field universality class, $\gamma = 1$,  and the scaling of long-range diluted lattices is recovered (see~\ref{subsec: Diluted long-range lattices}). In this mean-field region, the asymptotic behavior in Eq.~\eqref{eq: Delta_c C} only provides a good benchmark to the numerical data in the vicinity of the transition threshold, due to the strong fluctuations induced by the slow decaying disorder.  

On the other hand, for $4/3<\alpha<2$, the spectral dimension is $2< d_s<6$ and the percolation transition lies in the correlated regime with $\alpha$-dependent critical exponents~\cite{defenu2017percolation}. In particular, the inset of Fig.~\ref{fig: Squeezing_mxy_percolation}c shows the exponent $\gamma$, obtained via Monte Carlo simulations~\cite{defenu2017percolation}. In the correlated region $4/3<\alpha<2$ the disorder tails are weaker and the critical point is achieved for higher values of $C>0.25$, resulting in numerical data that closely follow the scaling in Eq.~\eqref{eq: Delta_c C}, see Fig.~\ref{fig: Squeezing_mxy_percolation}c. The numerical data for the scalable squeezing threshold (full circles in Fig.~\ref{fig: Squeezing_mxy_percolation}c with different colors for different values of $\alpha$) have been extracted by comparing numerical DTWA simulations (see App.~\ref{app: DTWA}) for different system sizes, $N=256,512,1024$, and identifying the threshold below which the (sufficiently long-time) minimum of the squeezing parameter~\eqref{eq: squeezing parameter} becomes size independent. The scaling in the $C\to C_p$ and $\Delta\to 1$ limit predicted in Eq.~\eqref{eq: Delta_c C} are shown as dashed lines in the plot. 

Figures~\ref{fig: Squeezing_mxy_percolation}d-f, compare the time evolution of the spin squeezing parameter $\xi^2$~\eqref{eq: squeezing parameter} and of the $xy$-magnetization $m_{xy}$ in the random graph modelling spatially correlated disorder, for varying system size $N$, interaction exponent $\alpha$, anisotropy $\Delta$, and dilution probability $p$, obtained via the DTWA method (see App.~\ref{app: DTWA}). Also in this case, once the data are interpreted in terms of the spectral dimension $d_s$, the scaling of the critical anisotropy near the Heisenberg point specific to the correlated disorder geometry~\eqref{eq: Delta_c C} is considered, and the role of the dilution probability is replaced by the strength of the power-law correlated disorder $C$, similar behavior is observed as for long-range diluted lattices and for the PW2 graph (Figs.~\ref{fig: Fig_examples}d–f). In particular, for $d_s>2$ ($\alpha<2$) and $\Delta>\Delta_c(C,\alpha)$ we observe long-time $xy$ order and scalable spin squeezing (Fig.~\ref{fig: Squeezing_mxy_percolation}d). For $d_s>2$ but $\Delta<\Delta_c(C,\alpha)$, no scalable spin squeezing is achieved (Fig.~\ref{fig: Squeezing_mxy_percolation}e). Finally, when $d_s<2$ ($\alpha>2$), spontaneous symmetry breaking is precluded, and scalable spin squeezing cannot be realized even in the absence of spatial disorder (Fig.~\ref{fig: Squeezing_mxy_percolation}f).

\section{Discussion and Outlook}\label{sec: Discussion}
Scalable quantum metrological gain is controlled by a subtle interplay between the structural properties and the interaction range of the system under study. In this work, we consider the possibility of achieving a squeezed state in the $xy$ plane of an XXZ ferromagnet, initialized in a semiclassical state fully polarized along $x$. The spin degrees of freedom sits on the nodes of an arbitrary network structure and interact via power-law decaying exchange interactions $\sim r^{-\alpha}$. By introducing an inhomogeneous spin-wave description of the model, we demonstrate that the emergence of OAT-like squeezing dynamics is inextricably tied with the scaling of the smallest (non-vanishing) gap of the Laplacian spectrum on the graph.

\begin{figure}%[h]
    \centering    \includegraphics[width=\linewidth]{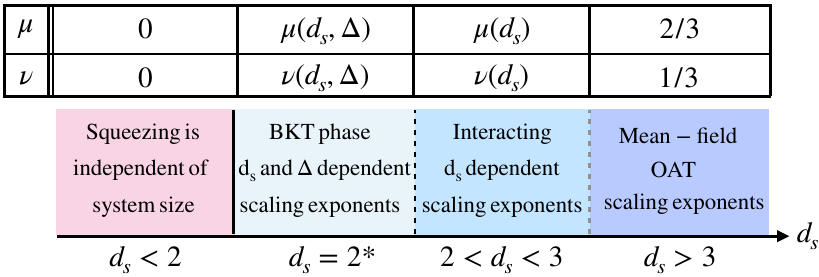}
     \caption{Schematic summary of the dynamical regimes of critical spin squeezing as a function of the spectral dimension $d_s$. For $d_s>3$, the dynamics belongs to the mean-field OAT universality class, with universal exponents independent of both $d_s$ and the anisotropy $\Delta$. In the intermediate regime $2<d_s<3$, scalable squeezing persists and is controlled by the finite-temperature critical point associated with xy-ferromagnetic order, leading to $d_s$-dependent critical exponents. At the marginal case $d_s=2^{*}$, scalable squeezing may survive through relaxation toward a Berezinskii–Kosterlitz–Thouless (BKT) phase, with nonuniversal scaling exponents continuously dependent on $\Delta$. Finally, for $d_s<2$, the absence of long-range order prevents scalable spin squeezing. $^*$The existence of a BKT phase in graph with spectral dimension $2$ is not guaranteed in general and it might depend on the specific network topology \cite{dyson1971anising, pagni2026onedimensional}.}
    \label{fig: Fig_SqueezingExponents}
\end{figure}
This allows us to establish sharp conditions under which the minimum squeezing parameter scales algebraically with the system size, $\xi^2_{\min}\sim N^{-\mu}$, with an optimal squeezing time $t_{\min}\sim N^\nu$. We demonstrate that the emergence of these scaling laws is  universal also on networks and is governed only by the spectral dimension of the graph, see Sec.~\ref{sec: Conditions for scalable spin-squeezing} for the details, and Fig.~\ref{fig: Fig_SqueezingExponents} for a schematic summary. (i) For $d_s>3$, the spin-squeezing dynamics is governed by the mean-field universality class of the ground-state quantum critical point crossed by the initial quench. In this regime, the dynamics reproduces the same universal mean-field scaling as the OAT model, with exponents $\mu = 2/3$ and $\nu = 1/3$, independently of the precise value of the spectral dimension or the anisotropy $\Delta$. (ii) For $2<d_s<3$, spin-wave fluctuations become relevant and the simple OAT description breaks down; nevertheless, scalable critical spin squeezing remains possible due to the existence of spontaneous symmetry breaking at finite temperature or energy density. In this interacting regime, the universal properties of the squeezing dynamics are governed by the finite-temperature critical point associated with the onset of $xy$-ferromagnetic order. Accordingly, the exponents $\mu$ and $\nu$ are expected to depend on the spectral dimension $d_s$, while remaining independent of the anisotropy $\Delta$. (iii) At the boundary value $d_s=2$, scalable squeezing could still survive if the system relaxes toward a Berezinskii–Kosterlitz–Thouless (BKT) phase, leading to an algebraic scaling of the squeezing parameter with a nonuniversal exponent continuously depending on the anisotropy $\Delta$ \cite{roscilde2024scalable}. However, the existence of a BKT phase is not guaranteed for generic XXZ or XY models defined on arbitrary graph geometries with spectral dimension $d_s=2$ \cite{dyson1971anising, pagni2026onedimensional}. (iv) Finally, for $d_s<2$, the absence of long-range order prevents any form of scalable spin squeezing in the thermodynamic limit. 

The extension and boundaries of the optimal squeezing region in the phase space $\alpha$ vs. $\Delta$ can then be quantified by means of a perturbative expansion in the weak anisotropy limit $\Delta\approx 1$ (Heisenberg limit). 

Extending this framework beyond the $|\Delta|<1$ regime represents a direction for future work. In particular, in the $\Delta\ll -1$ region, the rise of antiferromagnetic order offers a different mechanism to disrupt squeezing which still has to be explored. More generally, determining the universal scaling exponents of the squeezing parameter in the interacting regime ($2<d_s<3$) remains an open problem which we leave for future investigation.

Our results demonstrate that uncorrelated spatial disorder, either in the form of site or bond dilution, cannot fundamentally alter the spin-squeezing dynamics in the thermodynamic limit, despite causing substantial fluctuations at finite size as shown in the phase diagrams in Fig.~\ref{fig: Fig_examples}a and~\ref{fig: Fig_examples}b. Correlated disorder, on the other hand, alters the thermodynamic phase diagram yielding a subtle interplay between the dilution strength and the interaction decay.

A central outcome of our analysis is the existence of a critical anisotropy $\Delta_c$, whose scaling with system size and disorder strength is controlled by the competition between XXZ universality and percolation universality on the graph. As expressed in Eq.~\eqref{eq: Delta_c scaling}, this condition strongly constrains the possibility of achieving scalable squeezing in systems with low filling fractions or strong spatial disorder when the natural anisotropy is fixed by microscopic interactions. This limitation is particularly relevant for experimental platforms where $\Delta$ cannot be tuned continuously.

One possible route to overcome this constraint is provided by Floquet engineering techniques, which allow for an effective renormalization of the interaction anisotropy through periodic driving~\cite{kranzl2023observation,bukov2015universal}. By dynamically dressing the spin interactions~\cite{choi2020robust,zhou2020quantummetrology,martin2023controlling}, it is in principle possible to access regimes closer to the Heisenberg point $\Delta\to 1$, thereby enlarging the parameter space where scalable squeezing survives even in the presence of strong, possibly correlated, disorder. However, this strategy comes with an important trade-off: approaching the isotropic point leads to a parametric increase of the optimal squeezing time, which diverges as $t_{\min}\propto(1-\Delta)^{-1}$. As a consequence, longer coherence times are required, posing stringent constraints on experimental implementations.

An alternative and complementary strategy, consists in allowing the spins to move across the network using itinerant dipoles~\cite{bilitewski2021dynamical,wellnitz2024spin,douglas2025spin}. This scenario, spins initially occupying disconnected or non-percolating clusters can dynamically explore nearby sites, effectively mediating correlations between otherwise isolated regions of the graph. Such mobility-induced connectivity can partially restore collective dynamics and enable squeezing even below the static percolation threshold. Understanding the competition between interaction-driven squeezing and motional decoherence in this regime represents an interesting direction for future work.

Other promising directions for future research include extending our analysis beyond linear spin squeezing to more general forms of nonlinear squeezing that incorporate higher-order moments of the collective spin. Such extensions are known to capture richer forms of multipartite entanglement and can provide enhanced metrological gain beyond that detected by standard squeezing parameters, as previously demonstrated for OAT dynamics in terms of higher-order cumulants~\cite{gessner2019metrological}. In addition, the conventional notion of collective spin squeezing probes only uniform sensing directions, effectively restricting the metrological advantage to global observables. In contrast, inhomogeneous interaction networks can support nontrivial spatial structures in the correlations, leading to optimal sensing directions that are not aligned with collective spin components. These features can be systematically revealed by generalized spin-squeezing matrices, which account for mode-resolved fluctuations and are particularly relevant for multiparameter quantum sensing~\cite{gessner2020multiparameter}. Exploring these generalized notions of squeezing in complex network geometries may uncover new pathways to exploit spatially structured entanglement for enhanced and versatile quantum metrology.

More broadly, our results reveal that the interplay between $xy$-ferromagnetic interactions and strong spatial disorder gives rise to a rich and nontrivial universality structure, where properties traditionally associated with equilibrium critical phenomena on graphs directly impact the metrological performance. The emergence of disorder-dependent scaling laws for $\Delta_c$, $\xi^2_{\min}$, and $t_{\min}$ highlights the role of graph spectral properties as key organizing principles for collective quantum dynamics in inhomogeneous systems.  

Our predictions are well suited for experimental benchmarking in several state-of-the-art quantum simulators. Rydberg atom arrays offer a particularly promising platform to engineer nontrivial graph geometries, including diluted long-range lattices and hypergraph-like connectivity, while enabling direct access to spin squeezing and collective magnetization observables. Trapped-ion systems, on the other hand, provide control over power-law interactions in one and two dimensions, with tunable exponents in the range $0\lesssim\alpha\lesssim3$, allowing for a systematic exploration of the disorder–interaction interplay discussed here. 

%#######################################--ACKNOWLEDGMENTS--#######################################
\acknowledgments
We thank Christian Roos and Tommaso Roscilde for useful correspondance on the manuscript.
This research was funded by the Swiss National Science Foundation (SNSF) grant numbers 200021--207537 and 200021--236722, by the Deutsche Forschungsgemeinschaft (DFG, German Research Foundation) under Germany's Excellence Strategy EXC2181/1-390900948 (the Heidelberg STRUCTURES Excellence Cluster) and and the Swiss State Secretariat for Education, Research and Innovation (SERI). The authors acknowledge thee Wilczek Quantum Center in Shanghai for hosting the preliminar discussions during which the idea for this study was conceived. 

\section*{Data availability} 
The data and codes associated with this manuscript version are available under DOI:\href{https://doi.org/10.5281/zenodo.20330782}{10.5281/zenodo.20330782}

\section*{Note Added}
During the preparation of this work, two preprints~\cite{kaplanlipkin2025theoryscalablespinsqueezing,begg2026scalablespinsqueezingpowerlaw} appeared that investigate the effects of finite filling on spin squeezing in dipolar-interacting XXZ models on two-dimensional lattices. These systems correspond to a specific realization of the class of site-diluted long-range lattices analyzed in Sec.~\ref{subsec: Diluted long-range lattices}. When specialized to this setting, the predictions of our general theory are in full agreement with the conclusions of these works.  

In particular, Ref.~\cite{kaplanlipkin2025theoryscalablespinsqueezing} reports that the critical anisotropy $\Delta_c$ approaches the Heisenberg point in the limit of low filling (high dilution) with a square-root dependence on the filling fraction. This behavior is directly recovered from our general scaling relation in Eq.~\eqref{eq: Delta_c scaling} upon specifying to a two-dimensional site-diluted lattice with dipolar interactions. In this case, the system belongs to the mean-field percolation universality class with $p_p(N)\to 1$, mean cluster-size exponent $\gamma=1$, interaction exponent $\alpha=3$, and spatial dimension $d=2$.

For completeness, in App.~\ref{app: comparison} we extend the argument of Ref.~\cite{kaplanlipkin2025theoryscalablespinsqueezing} to generic power-law interactions and show that it leads to the same scaling behavior obtained by applying our general framework to diluted long-range lattices.
\appendix
\section{Spectral dimension of long-range lattices}\label{app: Spectral dimension of long-range lattices}
In this appendix we show the details of the spectrum low energy expansion for a long-range interacting system leading to the spectral dimension in Eq.~\eqref{eq: spectral dimension LR diluted}.

We consider the power law decaying coupligns
\begin{align}
    J_{l,j} = \frac{1}{\mathcal{N}_\alpha}\frac{1}{r_{l,j}^\alpha},
\end{align}
where $\mathcal{N}_\alpha = \sum_{r}r^{-\alpha}$ is a normalization factor necessary, for $\alpha<d$, to ensure energy extensivity~\cite{kac1963van}. More precisely, we notice that the Kac normalization $\mathcal{N}_\alpha$ scales differently with the system size $N\gg 1$ depending on $\alpha$:
\begin{align}
    \mathcal{N}_\alpha\approx \begin{cases}
        N^{d-\alpha} &\mathrm{if}\,\,\alpha<1\\
        \ln N &\mathrm{if}\,\,\alpha = 1\\
        \zeta(\alpha) &\mathrm{if}\,\,\alpha>1
    \end{cases}\,.\label{eq: kac scaling}
\end{align}

Leveraging the translational invariace of the coupling matrix it is convinient to introduce it Fourier transform 
\begin{align}
    \tilde{J}_k(\alpha) = \frac{1}{\mathcal{N}_\alpha}\sum_{r = 1}^{N}\frac{\cos(kr)}{r^{\alpha}},\label{eq: falpha}
\end{align}
Assuming periodic boundary conditions, we have the usual restriction on the momentum $k\equiv k_n = 2\pi n/N$ with $m\in \mathbb{Z}$ and $m = \lfloor -N/2\rfloor,\dots\lfloor N/2\rfloor$ (the lattice spacing has been set to 1). The spectral gap, setting the spectral dimension, is determined by the minimum gap with respect to the zero mode
\begin{align}
\delta\lambda =\min_{k\neq 0}[\tilde{J}_0(\alpha)-\tilde{J}_k(\alpha)]\approx N^{-2/d_s}\,.
\end{align}
If the minimum is attained by the second lowest lying Fourier mode $k_1 = 2\pi/N$ and $N\to\infty$ limit corresponds to a well defined continnuous limit in the $k$ space. It follows that the spectral dimension can also be determined by the dispersion relation close to $k\approx 0$ using the relation
\begin{align}
\tilde{J}_0(\alpha)-\tilde{J}_k(\alpha)\approx |k|^{2d/d_s}\,.\label{eq: dispersion ds}
\end{align} 

\subsection{Weak long-range ($\alpha>d$)}
As long as we are in the weak long-range regime $\alpha>d$, the Kac scaling is finite in the $N\to\infty$ limit. Accordingly the calculation proceeds similarly to the nearest-neighbor case, allowing the thermodynamic limit of Eq.~\eqref{eq: falpha} to be taken safely, substituting the discrete momentum values $k_n$ with the continuous variable $k\in[-\pi,\pi)$. 

Let us start, for simplicity, from the one dimensional case $d = 1$. Then, for $\alpha>1$ and in the $N\to\infty$ limit the couplings Fourier transform reads~\cite{defenu2019universal}
\begin{align}
    \tilde{J}_k(\alpha) \approx \frac{1}{2\zeta(\alpha)}\left[\mathrm{Li}_\alpha(e^{ik})+\mathrm{Li}_\alpha(e^{-ik})\right]\label{eq: falpha weak LR}
\end{align}
where $\mathrm{Li}_x(z) = \sum_{n = 1}^{\infty}z^n/n^x$ is the polylogarithm and and $\zeta(x)$ is the Riemann zeta function~\cite{abramowitz1964handbook}.

In order to determine the spectral dimension $d_s$, we are interested in the low $k$ modes of the single particle spectrum, which determines the dispersion relation.

This is obtained by taking the Taylor expansion of Eq.~\ref{eq: falpha weak LR} around $k = 0$ leading to~\cite{defenu2019universal}
\begin{align}
\tilde{J}_k &= 1+\sin\left(\frac{\alpha\pi}{2}\right)\frac{\Gamma(1-\alpha)}{\zeta(\alpha)}|k|^{\alpha-1}+\mathcal{O}(|k|^2),
\end{align}
for $1<\alpha<3$,
\begin{align}
\tilde{J}_k &=  1+\frac{2\ln(k)-3}{4\zeta(3)}|k|^2 + \mathcal{O}(|k|^3),
\end{align}
for $\alpha = 3$,
\begin{align}
\tilde{J}_k &=  1-\frac{\zeta(\alpha-2)}{2\zeta(\alpha)}|k|^2 + \mathcal{O}(|k|^{\alpha-d}),
\end{align}
for $\alpha > 3$.

In order to generalize the above results to the $d>1$ case we follow the standard procedure for Fourier transforming a radial function, we switch to spherical coordinates and integrate over all the angles obtaining~\cite{defenu2024outofequilibrium}
\begin{align}
    \tilde{J}_{\boldsymbol{k}}\approx \frac{2^{d/2-1}\Gamma(d/2)\int_{1}^{\infty} d\rho\rho^{d-1-\alpha}\mathcal{J}_{d/2-1}(|\boldsymbol{k}|\rho)(|\boldsymbol{k}|\rho)^{1-d/2}}{\int_1^\infty d\rho\rho^{d-1-\alpha}},\label{eq: Fourier d}
\end{align}
where $\mathcal{J}_\nu(x)$ is the standard Bessel function of order $\nu$~\cite{abramowitz1964handbook}. Expanding Eq.~\eqref{eq: Fourier d} in the $|\boldsymbol{k}|\approx 0$ limit we obtain
\begin{align}
\tilde{J}_{\boldsymbol{k}}&\approx 1-A(\alpha)|\boldsymbol{k}|^{\alpha-d}+B(\alpha)|\boldsymbol{k}|^2+\mathcal{O}(|\boldsymbol{k}|^4)
\end{align}
where 
\begin{align}
    A(\alpha) = \frac{\Gamma(d/2)\Gamma((2-\alpha+d)/2)}{2^{\alpha-d}\Gamma(\alpha/2)}\,,
    \end{align}
and 
\begin{align}
    B(\alpha) = \frac{\alpha-d}{2d^2+4d-2\alpha d}\,.
\end{align}
The first (second) term governs the asymptotic low-momentum behavior of $\tilde{J}_{\boldsymbol{k}}$ for $d<\alpha<d+2$ (respectively $\alpha>d + 2$), leading to the dispersion relation
\begin{align}
   \tilde{J}_0(\alpha)-\tilde{J}_k(\alpha)\approx\begin{cases}
       |\boldsymbol{k}|^{\alpha-d} &d<\alpha<d+2\\
       |\boldsymbol{k}|^{2} &\alpha>d+2
   \end{cases}\,.
\end{align}
Finally comparing this result with Eq.~\eqref{eq: dispersion ds} we obtain the spectral dimension for long-range lattices
\begin{align}
    d_s = \begin{cases}
       2d/(\alpha-d)  &d<\alpha<d+2\\
       d              &\alpha>d+2
   \end{cases}\,,
\end{align}
corresponding to Eq.~\eqref{eq: spectral dimension LR diluted} of the main text.
\subsection{Strong long-range ($0<\alpha<d$)}\label{subapp: Strong long-range}
The situation changes dramatically in the strong long-range regime $\alpha < d$. Indeed, as shown in Eq.~\eqref{eq: kac scaling}, the Kac normalization factor $\mathcal{N}_\alpha$ diverges at large $N$ ensuring energy extensivity. Accordingly, the thermodynamic limit of Eq.~\eqref{eq: falpha} must be carefully considered. To this aim, it is convenient to write Eq.~\eqref{eq: falpha} explicitly for large $N$ as 
\begin{align}
    \lim_{N\to\infty}\frac{1}{\mathcal{N}_\alpha}\sum_{r = 1}^{N/2-1}\frac{\cos(kr)}{r^\alpha}\approx\frac{c_\alpha}{N}\sum_{r = 1}^{N/2}\frac{\cos(2\pi n \frac{r}{N})}{(r/N)^\alpha}.
\end{align}
Due to the $1/N$ scaling of the discrete momenta on the lattice, the summation depends only on the variable $r /N$. Therefore, for $N\to\infty$, we can take the continuum limit by transforming the sum over $r$ into an integral with respect to $s = r/N$, leading to ~\cite{defenu2021metastability}
\begin{align}
    \tilde{J}_n(\alpha) = \lim_{N\to\infty}\tilde{J}_k(\alpha) = c_\alpha\int_0^{1/2} ds\frac{\cos(2\pi n s)}{s^\alpha}.\label{eq: falpha spectrum}
\end{align}
As a consequence the spectrum remains discrete even as $N\to\infty$. Specifically, for $\alpha<d$ , the gap between neighboring eigenvalues $\lambda_{n+1}-\lambda_n$, labeled by the consecutive momenta $k_n$, $k_{n+1}$ in Eq.~\eqref{eq: falpha}, does not vanish in the thermodynamic limit, as it would for $\alpha>d$. Therefore, the eigenvalues depend only on the integer index $n\in \mathbb{Z}$  rather than on the continuous momentum $k$:
\begin{align}
    \lambda_n = \tilde{J}_0(\alpha) - \tilde{J}_n(\alpha).\label{eq: discrete spectrum}
\end{align}
Notably, for $\alpha = 0$, we find that  $\tilde{J}_n(\alpha)\to\delta_{n,0}$, leading to a fully degenerate discrete spectrum as described by Eq.~\eqref{eq: discrete spectrum}: $\lambda_n = \tilde{J}_0(\alpha)$ for $n\neq 0$ and $\lambda_n = 0$ for $n=0$. Additionally, we observe that the eigenvalues $\lambda_n$ are not densely distributed. Instead, each eigenvalue is isolated, with the only accumulation point occurring at the maximum $\max_n\lambda_n = \tilde{J}_0(\alpha)$. This follows from the Riemann–Lebesgue lemma~\cite{last1996quantum}, which implies 
\begin{align}
\lim_{n\to\infty}\tilde{J}_n(\alpha) = 0.
\end{align}

Finally, since the spectral gap $\delta\lambda$ remains finite in the thermodynamic limit, and more precisely equal to
\begin{align}
    \delta\lambda\approx 1-c_\alpha\int_0^{1/2} ds\frac{\cos(2\pi s)}{s^\alpha},
\end{align}
it follows that the spectral dimension of long-range lattices in the strong long-range regime $0<\alpha<d$ is always infinite 
\begin{align}
    d_s = \infty\quad\mathrm{if}\quad0<\alpha<d\,.
\end{align}
\subsection{Role of an exponential cutoff}\label{app: Role of an exponential cutoff}
We analyze the effect of introducing an exponential cutoff in the interaction profile, replacing the pure power-law couplings with
\begin{align}
J_r \sim r^{-\alpha} e^{-\kappa r}.
\end{align}
The Fourier transform of the couplings then reads 
\begin{align}
    \tilde{J_k}(\alpha) = \sum_{r=1}^N\frac{\cos(kr)e^{-\kappa r}}{r^{\alpha}},
\end{align}
and the spectral gap between the zero mode and the first Fourier mode $k_1 = 2\pi/N$ is
\begin{align}
\delta\lambda
= \tilde{J}_0(\alpha) - \tilde{J}_{k_1}(\alpha)
= \sum_{r=1}^{N} \frac{\left[1-\cos(k_1 r)\right] e^{-\kappa r}}{r^{\alpha}}.
\end{align} 

In several experimental settings, the exponential cutoff scales with system size as $\kappa = \Lambda/N$, where $\Lambda=\mathcal{O}(1)$~\cite{schuckert2025observation}. Making the $N$ dependence explicit through $k_1=2\pi/N$ and $\kappa=\Lambda/N$, we obtain
\begin{align}
    \delta\lambda =\frac{1}{N^\alpha}\sum_{r=1}^N\frac{[1-\cos(2\pi (r/N))]e^{-\Lambda (r/N)}}{(r/N)^{\alpha}}\,.
\end{align}
The $N\to\infty$ limit then corresponds to a continuum limit with respect to the variable $x=r/N$ allowing to pass from a discrete sum other $r$ to an integral over $x$
\begin{align}
    \delta\lambda \approx N^{1-\alpha}\int_{1/N}^1dx\frac{[1-\cos(2\pi x)]e^{-\Lambda x}}{x^{\alpha}}\,.
\end{align}
The convergence of the integral as $N\to\infty$ is controlled by the integrand behavior as $x\sim1/N\to0$
\begin{align}
    \frac{[1-\cos(2\pi x)]e^{-\Lambda x}}{x^{\alpha}}\sim x^{2-\alpha}\,.
\end{align}
Similarly to the case without any exponential cutoff, this allows to identify two regimes: if $0<\alpha<2$ the integral converges in the thermodynamic limit, yielding $\delta\lambda\sim N^{1-\alpha}$; if $\alpha>2$, the integral diverges as $\sim N^{\alpha-3}$, leading to $\delta\lambda\sim N^{-2}$. 

Therefore, the presence of an exponential cutoff does not modify the scaling of the spectral gap, provided that the decay rate scales as $\kappa \sim 1/N$ with the system size. Consequently, the spectral dimension remains unchanged with respect to the pure power-law case.
\section{Spectral dimension of the power-of-two graph}
\label{app: Expansion of the power-of-two graph spectrum}

In this Appendix we analyze the low-energy structure of the spectrum of the power-of-two (PW2) graph introduced in Sec.~\ref{subsec: Hypergraphs structures}. The PW2 graph is defined on a one-dimensional ring of $N$ sites, with couplings connecting only pairs of sites whose distance is an integer power of two. Explicitly, the couplings read
\begin{align}
J_{i,j} =
\begin{cases}
|i-j|^{-\alpha}, & \text{if } |i-j| = 2^{n},\\
0, & \text{otherwise}.
\end{cases}
\end{align}
with $n=0,1,\dots,\log_2(N)-1$.

In this case, the scaling of the Kac normalization  then reads
\begin{align}
\mathcal{N}_{\alpha,\mathrm{PW2}} = \sum_{r=0}^{\log_2(N)-1} 2^{-\alpha r}.
\end{align}
This geometric series can be summed explicitly, yielding in the large-$N$ limit three distinct regimes:
\begin{align}
\mathcal{N}_{\alpha,\mathrm{PW2}} \approx
\begin{cases}
\dfrac{N^{-\alpha}}{2^{\alpha}-1}, & \alpha < 0,\\[6pt]
\log_2(N), & \alpha = 0,\\[6pt]
\dfrac{1}{2^{\alpha}-1}, & \alpha > 0.
\end{cases}
\end{align}
For $\alpha<0$, the interaction strength increases with distance, leading to a divergence of the effective coupling with system size, therefore we will need to introduce a normalization factor scaling as $N^{\alpha}$ to cure this divergence ensuring energy extensivity and a meaningful thermodynamic limit also in this case. 

In the absence of spatial disorder, translational invariance allows us to diagonalize the coupling matrix in momentum space. Introducing the discrete Fourier transform, the spectrum reads
\begin{align}
\tilde{J}_k(\alpha)
=
\sum_{n=0}^{\log_2(N)-1}
2^{-\alpha n}\cos\!\left(k\,2^n\right),
\label{eq: spectrum PW2}
\end{align}
where $k=2\pi m/N$ with $m = \lfloor -N/2\rfloor,\dots\lfloor N/2\rfloor$, assuming periodic boundary conditions. Also in this case the spectral gap is determined by the minimum gap with respect to the zero mode
\begin{align}
\delta\lambda =\min_{k\neq 0}[\tilde{J}_0(\alpha)-\tilde{J}_k(\alpha)].
\end{align}
\subsection{The $\alpha = 0$ case}
For $\alpha=0$, all nonzero couplings have equal strength. Equation~\eqref{eq: spectrum PW2} reduces to a binary sum of cosine functions. The minimal gap is attained at momentum $k=\pi$, for which
\begin{align}
\tilde{J}_0 - \tilde{J}_\pi
&=
\sum_{n=0}^{\log_2(N)-1}
\left[1-\cos(2^n\pi)\right]=2,
\end{align}
independently of system size. Thus, similarly to strongly long-range interacting lattices with $\alpha<d$ (see App.~\ref{subapp: Strong long-range}), the spectrum of the PW2 graph at $\alpha=0$ remains gapped in the thermodynamic limit~\cite{kuriyattil2025entangled}. As a consequence, the effective spectral dimension diverges, $d_s\to\infty$.

\subsection{The $\alpha>0$ case}
For $\alpha>0$, the sum in Eq.~\eqref{eq: spectrum PW2} converges as $N\to\infty$ and defines a classical Weierstrass function~\cite{weierstrass1988,hardy1916weierstrass},
\begin{align}
W(x) = \sum_{n=0}^{\infty} \lambda^{n}\cos(x b^{n}),
\end{align}
with parameters
\begin{align}
\lambda = 2^{-\alpha}, \qquad b=2,
\end{align}
satisfying $0<\lambda<1$ and $b>1+3\pi/2$. The Weierstrass function is continuous everywhere but nowhere differentiable, and exhibits a self-similar fractal structure.

To extract the effective dispersion at small momenta, we first work at finite $N$ and expand the cosine in Eq.~\eqref{eq: spectrum PW2} for small $k$:
\begin{align}
    \tilde{J}_0-\tilde{J}_k &\approx\frac{k^2}{2}
\sum_{n=0}^{\log_2(N)-1}
2^{n(2-\alpha)}
\nonumber\\
&=\frac{k^2}{2}\frac{1-N^{2-\alpha}}{1-2^{2-\alpha}}\,.
\end{align}
Then to obtain the correct power law behavior as $N\to\infty$, one must take into account that the smallest nonzero momentum scales as $k_1=2\pi/N$. Sending $N\to\infty$ while keeping $kN=\mathcal{O}(1)$, two distinct regimes emerge
\begin{align}
\tilde{J}_0 - \tilde{J}_k \approx
\begin{cases}
k^2 & \alpha > 2, \\
k^{\alpha} & 0 < \alpha < 2.
\end{cases}\label{eq: PW2 dispersion app}
\end{align}
This leads to the spectral dimension
\begin{align}
d_s =
\begin{cases}
2 & \alpha > 2, \\
2/\alpha & 0 < \alpha < 2.
\end{cases}
\end{align}

A more formal derivation follows from the rigorous bounds on the Weierstrass function~\cite{david2018bypassing}
\begin{align}
    C_{\inf}|x-x'|^{2-d_H}\leq |W(x)-W(x')|\leq C_{\sup}|x-x'|^{2-d_H},
\end{align}
where $C_{\inf}$ and $C_{\sup}$ are strictly positive constants and $d_H$ denotes the Hausdorff (fractal) dimension. For the Weierstrass function, this dimension is known to be related to the $\lambda$ and $b$ parameters as~\cite{hunt1998thehausdorff,david2018bypassing}
\begin{align}
    d_H = 2+\frac{\ln\lambda}{\ln b} = 2-\alpha,\label{eq: PW2 Hausdorff dimension}
\end{align}
where in the last equality we used the specific parameters of the PW2 graph. Choosing $x=0$ and $x'=k$ and taking $k\to 0$ immediately reproduces the dispersion scaling in Eq.~\eqref{eq: PW2 dispersion app}. This also yields a direct relation between spectral and fractal dimensions
\begin{align}
    d_s = 2(2-d_H)^{-1},
\end{align}
which provides a general framework for defining the spectral dimension of graph spectra generated by Weierstrass functions with arbitrary $\lambda$ and $b$.

\subsection{The $\alpha<0$ case}
For $\alpha<0$, extensivity requires the introduction of a normalization factor $N^{\alpha}$, leading to the rescaled spectrum
\begin{align}
\tilde{J}_k(\alpha)
&=
\sum_{n=0}^{\log_2(N)-1}
N^\alpha2^{-\alpha n}\cos\left(k\,2^n\right)\notag\\
&=\sum_{n=0}^{\log_2(N)-1} 2^{\alpha\left[\log_2(N)-n\right]}\cos\left(2^n k\right).
\label{eq: spectrum PW2 negatrive alpha}
\end{align}
Because $\alpha<0$, the dominant contributions now come from large distances $n\sim\log_2(N)$ rather than from short-range terms. This becomes explicit upon performing the change of variables $n'=\log_2(N)-1-n$ (related to the so called Monna map~\cite{monna1952transformation}), yielding
\begin{align}
\tilde{J}_k(\alpha) &= 2^{\alpha}\sum_{n=0}^{\log_2(N)-1} 2^{\alpha n}\cos\left(2^{-n}\tilde{k}\right)\,,
\end{align}
where we introduced the rescaled momenta $\tilde{k}=Nk/2$. Now recalling that $k=k_m = 2\pi m/N$, with $m$ and integer such that $m = -N/2,\dots,N/2$ (assuming periodic boundary conditions), we notice that the first mode after the zero mode $k_0 = 0$, is no more given by $k_1= 2\pi/N$. Indeed, in the sum
for $m=1$ we have that $\tilde{k}_1 = Nk_1/2 = \pi$, meaning that this mode remains gapped in the thermodyanic limit $N\to\infty$. Instead, similarly to the $\alpha = 0$ case, the minimal gap corresponds to the $k=\pi$ mode. The corresponding spectral gap is given by
\begin{align}
    \delta\lambda &= J_0(\alpha)-J_\pi(\alpha) \\&=N^\alpha\sum_{n=0}^{\log_2(N)-1}2^{-\alpha n}[1-\cos\left(\pi\,2^n\right)]\,.
\end{align}
Since $\cos(\pi,2^n)=1$ for all $n\geq1$, only the $n=0$ term contributes, yielding
\begin{align}
\delta\lambda = N^{\alpha} = N^{-|\alpha|}.
\end{align}
This scaling implies a spectral dimension
\begin{align}
d_s = \frac{2}{|\alpha|},
\end{align}
in agreement with Eq.~\eqref{eq: ds PW2}.
\begin{figure*}
    \centering    \includegraphics[width=\linewidth]{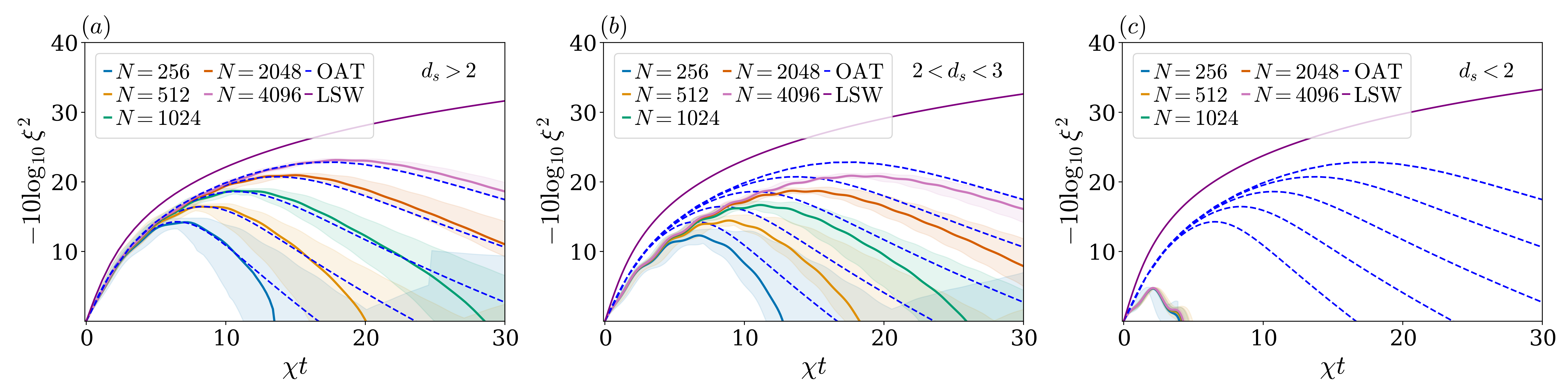}
   \caption{Spin-squeezing dynamics, expressed as $-10\log_{10}\xi^2$, for a one-dimensional long-range diluted lattice. Solid lines show the rotor/spin-wave prediction Eq.~\eqref{eq: xi RotoSW}, averaged over $400$ disorder realizations, for different system sizes $N$ (color-coded curves). Panels correspond to different interaction exponents: (a) $\alpha = 1.2$ ($d_s=10>3$), (b) $\alpha = 1.8$ ($2<d_s=2.5<3$), and (c) $\alpha = 2.8$ ($d_s\approx 1.11<2$). The dilution probability is $p = 0.2$ and the anisotropy is $\Delta=0$. Shaded regions represent individual disorder realizations. Blue dashed lines denote the reference OAT dynamics for an effective system size $(1-p)N$.Purple solid curves correspond to the linear spin-wave approximation, which neglects the nonlinear rotor dynamics.}\label{fig: Rotor_SW}
\end{figure*}
\section{Additional details on the Rotor/Spin-Wave theory results}\label{app: Rotor-Spin wave}
In this Appendix we provide additional details on the rotor/spin-wave theory results discussed in Sec.~\ref{sec: Rotor/Spin-wave theory on graphs}. Within this approximation, the dynamics of the spin-squeezing parameter,\eqref{eq: squeezing parameter} can be computed analytically by separating the collective (zero-mode) contribution from the finite-momentum spin-wave excitations. 

We consider
\begin{align}
    \xi_R^2 = \frac{N\min_{\perp}[\mathrm{Var}(S_{\perp})]}{\langle S_x\rangle^2},
\end{align}
where $S_{\perp} =  \cos\theta S_y + \sin \theta S_z$, so that
\begin{align}
    \mathrm{Var}(S_{\perp}) &= \cos^2\theta\mathrm{Var}(S_{y}) + \sin^2\theta\mathrm{Var}(S_{z})\notag\\
    &+2\sin\theta\cos\theta\mathrm{Cov}(S_y,S_z),
\end{align}
where $\mathrm{Var}(S_{a}) =  \langle S_a^2\rangle-\langle S_a\rangle^2$ and $\mathrm{Cov}(S_a,S_b) = \frac{1}{2}\langle\{ S_a,S_b\}\rangle- \langle S_a\rangle\langle S_b\rangle$, with $a = x,y,z$, are the variance and covariance of the total spin operators, respectively.

Since our spin-wave expansion is performed around an $x$-magnetized state then the expectation value of the $x$ component of the total spin $\mathbf{S}$ is given by 
\begin{align}
    \langle S_x\rangle = Ns-\sum_n\langle a_n^\dagger a_n\rangle,
\end{align}
which, within the rotor/spin-wave decomposition, becomes~\cite{roscilde2023rotor}
\begin{align}
    \langle S_x\rangle = \langle K_x\rangle_{\rm R}-\sum_{n\neq 0}\langle a_n^\dagger a_n\rangle_{\rm SW}.
\end{align}
Here, $\langle\cdot\rangle_{\rm R}$ denotes expectation values in the collective rotor mode, while $\langle\cdot\rangle_{\rm SW}$ refers to the finite-momentum spin-wave sector.

Moreover, as shown in Ref.~\cite{roscilde2023rotor}, the transverse fluctuations are dominated by the rotor contribution leading to
\begin{align}
    &\mathrm{Var}(S_y)\approx\mathrm{Var}(K_y)_{\rm R},\\
    &\mathrm{Var}(S_z)\approx\mathrm{Var}(K_z)_{\rm R},\\
    &\mathrm{Cov}(S_y,S_z)\approx \frac{1}{2}\langle\{K_y, K_z\}\rangle_{\rm R}\,.   
\end{align}
Therefore, the squeezing parameter takes the form
\begin{align}
    \xi_R^2\approx \frac{N \min_{\perp}[\mathrm{Var}(K_\perp)_{\rm R}]}{\langle K_x\rangle_{\rm R}-\sum_{n\neq 0}\langle a_n^\dagger a_n\rangle_{\rm SW}}.
\end{align}
The rotor sector can be solved exactly and reproduces the OAT result~\cite{kitagawa1993squeezed}
\begin{align}
    \langle K_x(t)\rangle_R &= \frac{N}{2}\cos^{N-1}(\chi t)\\
    \min_{\perp}[\mathrm{Var}(K_\perp(t))_{\rm R}] &=\frac{N}{4}+\frac{N(N-1)}{16}\left(A+\sqrt{A^2+B^2}\right), 
\end{align}
where 
\begin{align}
    A=1-\cos^{N-2}(2\chi t) ,\,
    B =  4\sin(\chi t)\cos^{N-2}(\chi t),
\end{align}
and, in our case, the rotor frequency $\chi$ is related to the average degree of the graph as
\begin{align}
    \chi\approx\frac{1}{2I}\approx \mathrm{deg}\frac{1-\Delta}{2(N-1)}\,.
\end{align}

The finite-momentum modes are governed by a quadratic bosonic Hamiltonian and can be solved analytically in the Laplacian eigenbasis. Defining the correlation functions $G_n = \langle a_n^\dagger a_n\rangle$ and $F_n = \langle a_n a_n\rangle$, one obtains
\begin{align}
    G_n(t) &= 2 \mathcal{U}_n^2 \mathcal{V}_n^2[1-\cos(2\omega_nt) ]\\
    F_n(t) &= \mathcal{U}_n \mathcal{V}_n(\mathcal{U}_n^2e^{-2i\omega_n t}+\mathcal{V}_n^2e^{2i\omega_nt}-2\mathcal{V}_n^2-1),
\end{align}
where $\omega_n$ are the spin-wave frequencies and $\mathcal{U}_n,\mathcal{V}_n$ are the Bogoliubov coefficients.

Combining the two contributions, the squeezing parameter reads
\begin{align}
\xi_R^2\approx\frac{\left(A+\sqrt{A^2+B^2}\right)}{4\left(\cos^{N-1}(\chi t)-\frac{2}{N}\sum_{n\neq 0}G_n(t)\right)^2}\,.\label{eq: xi RotoSW}
\end{align}

Figure~\ref{fig: Rotor_SW} shows the dynamics obtained from Eq.~\eqref{eq: xi RotoSW} for one-dimensional long-range diluted lattices at different system sizes and interaction exponents $\alpha$, corresponding to different spectral dimensions $d_s$.

A first benchmark is provided by linear spin-wave (LSW) theory (purple curves), where the zero mode is also treated as a non-interacting bosonic degree of freedom. In this approximation, the initial state evolves as a freely expanding Gaussian wavepacket, leading to the squeezing of the collective-spin component in the $yz$-plane~\cite{roscilde2023rotor,roscilde2023entangling}. However, the occupation of the zero mode grows without bound, resulting in an indefinitely decreasing squeezing parameter. This behavior neglects the compact nature of the collective spin and fails to capture finite-size effects, as well as the dependence on $\alpha$ and dilution $p$. Consequently, LSW theory breaks down at times of order $\mathcal{O}(1)$.
In contrast, the rotor/spin-wave approach properly accounts for the nonlinear dynamics of the collective mode, regularizing the growth of fluctuations and yielding a well-defined minimum of $\xi^2$. This framework thus provides a controlled description of the intermediate-time dynamics where optimal squeezing is generated, as well as its dependence on the underlying graph structure through the spectral properties of the Laplacian. In particular, for $d_s>3$ (Fig.~\ref{fig: Rotor_SW}a), the disorder-averaged squeezing dynamics closely follows the OAT result for an effective system size $(1-p)N$, where $p$ is the dilution probability, as shown by the blue dashed lines. For $2<d_s<3$ (Fig.~\ref{fig: Rotor_SW}b), the role of the spin wave excitations on top of the zero mode becomes relevant before the time scale set by the rotor spectral gap, leading to visible deviations from the ideal OAT behavior while still preserving scalable optimal squeezing. Finally, for $d_s<2$ (Fig.~\ref{fig: Rotor_SW}c), no scalable squeezing is observed: after an initial, non-scalable minimum at short times, spin-wave excitations proliferate and the approximation itself breaks down, resulting in an unbounded decrease of $\xi^2$ (similar to LSW), followed at later times by a sharp increase associated with the decay of the collective magnetization (not shown). The shaded regions in Fig.~\ref{fig: Rotor_SW} represent individual disorder realizations; their spread decreases with increasing system size, highlighting the self-averaging nature of the spin-squeezing dynamics in the thermodynamic limit.
\section{Discrete truncated Wigner approximation}\label{app: DTWA}
In this Appendix we provide additional details on the numerical method used to simulate the spin-squeezing dynamics in the main text. Our simulations are based on the Discrete Truncated Wigner Approximation (DTWA), a semiclassical phase-space method that allows one to approximate the real-time dynamics of large spin systems by sampling an ensemble of classical trajectories ~\cite{schachenmayer2015manybody,zhu2019generalized}.

DTWA maps the quantum dynamics of spin-$1/2$ operators onto an ensemble of classical spin variables evolving under mean-field equations of motion. For each spin $i$, we introduce a classical vector $\mathbf{s}  = (s^x_i,s^y_i,s^z_i)^T$ whose initial values are sampled from a discrete Wigner distribution reproducing the quantum expectation values of the initial state. For the fully $x$-polarized initial state the DTWA prescription assigns deterministic initial conditions along the polarization axis and stochastic transverse fluctuations
\begin{align}
s_i^x(0) = s,\quad
s_i^y(0) = \pm s,\quad
s_i^z(0) = \pm s,
\end{align}
where $s$ is the spin length (in all our examples $s=1/2$) and the signs of $s_i^y(0)$ and $s_i^z(0)$ are chosen independently with equal probability.
Each sampled configuration is then evolved according to classical mean-field equations of motion derived from the Heisenberg equations,
\begin{align}
\frac{d\mathbf{s}_i}{dt} = \mathbf{s}_i
\times
\mathbf{B}_i^{\mathrm{eff}},
\end{align}
where the effective field acting on spin $i$ is
\begin{align}
B_i^x &= \sum_j J_{ij}s_j^x,\\
B_i^y &= \sum_j J_{ij}s_j^y,\\
B_i^z &= \Delta \sum_j J_{ij}s_j^z.
\end{align}
These equations conserve the classical spin length $|\mathbf{s}_i|=s$ and correspond to a nonlinear system of $3N$ coupled ordinary differential equations. Expectation values of observables are obtained by averaging over many stochastic trajectories
\begin{align}
    \langle\hat{O}(t)\rangle\approx \frac{1}{\mathcal{N}_\mathrm{s}}\sum_{l}O_\mathrm{cl}^{(l)}(t)= \langle O_\mathrm{cl}^{(l)}\rangle_s\,,
\end{align}
where $\mathcal{N}_\mathrm{s}$ is the number of samples. In the disordered case we perform the sampling over DTWA trajectories and over the disorder at the same time and we use $\mathcal{N}_\mathrm{s} = 500$ samples. For each stochastic realization we generate a coupling matrix $J_{ij}$ corresponding to the graph under consideration and integrate the equations of motion using an adaptive Runge–Kutta solver.

From each trajectory we compute the collective spin
\begin{align}
\mathbf{S}(t)=\sum_{i=1}^N \mathbf{s}_i(t),
\end{align}
and accumulate ensemble averages of its first and second moments. Denoting by $\langle\dots\rangle_s$ the stochastic average, we evaluate the transverse variances
\begin{align}
\mathrm{Var}(S_y)
&=
\langle S_y^2\rangle_s-\langle S_y\rangle_s^2,\\
\mathrm{Var}(S_z)
&=
\langle S_z^2\rangle_s-\langle S_z\rangle_s^2,
\end{align}
and the covariance
\begin{align}
\mathrm{Cov}(S_y,S_z)
=\langle S_y S_z\rangle_s
-\langle S_y\rangle_s\langle S_z\rangle_s.
\end{align}
The minimal spin fluctuation in the $yz$ plane is given by
\begin{align}
&\Delta S_\perp^2
= \frac{1}{2}[\mathrm{Var}(S_y)+\mathrm{Var}(S_z)]\notag\\
&
-\frac{1}{2}\sqrt{
(\mathrm{Var}(S_z)-\mathrm{Var}(S_y))^2
+
4\mathrm{Cov}(S_y,S_z)^2
}\,.
\end{align}
The spin-squeezing parameter is then computed as
\begin{align}
\xi^2(t)
=\frac{N,\Delta S_\perp^2(t)}
{\langle S_x(t)\rangle^2}\,.
\end{align}
Moreover, we compute the average magnetization in the $xy$-plane as
\begin{align}
    m_{xy} = \frac{1}{N}\sqrt{\langle S_x^2\rangle_s+\langle S_y^2\rangle_s}\,.
\end{align}

In the disorder-free case DTWA has been validated against other numerical methods for the dynamics of the XXZ model with power law decaying interactions and it has been shown to yield nearly exact results~\cite{muleady2023validating}. It is widely used to capture the scaling behavior of the spin squeezing dynamics~\cite{schachenmayer2015manybody,perlin2020spin, block2024scalable}. Moreover, it has been used to treat several systems in presence of disorder~\cite{acevedo2017exploring,covey2018approach,kelly2021effect,signoles2021glassy,schultzen2022semiclassical}.

\section{Comparison with Ref.~\cite{kaplanlipkin2025theoryscalablespinsqueezing}}\label{app: comparison}
For completeness, in this Appendix we show that the general scaling relation for the critical anisotropy close to the percolation transition derived in Eq.~\eqref{eq: Delta_c scaling} is consistent, when specialized to site-diluted long-range lattices, with the scaling obtained by generalizing the argument of Ref.~\cite{kaplanlipkin2025theoryscalablespinsqueezing}. In that work, the authors analyzed two dimensional dipolar systems with finite filling fraction; here we extend their reasoning to generic $d$ dimensional lattices with power-law interactions decaying as $r^{-\alpha}$.  

The idea from~\cite{kaplanlipkin2025theoryscalablespinsqueezing} to find the scaling of the critical point at low filling (close to the percolation transition) is to consider the contribution to the magnetic order coming from isolated dimers (clusters of size $\sim 2$) of spins. In our language this translates to the fact that while below the percolation transition the system breaks up into disconnected components of size $\sim 1$, just above the transition, $p\lesssim p_p(N)$, the most probable connected clusters are dimers, which therefore provide the dominant contribution to the onset of magnetic order.

At strong dilution, the average energy associated to a single dimer is proportional to the anisotropy $(1-\Delta)$ while the density of spins is $\rho = (1-p)/a^d$, where $a$ is the lattice spacing of the underlying clean lattice. Accordingly, the average energy scale associated with dimers can be estimated as
\begin{align}
\overline{\varepsilon}_\mathrm{dimers}\sim \frac{(1-p)(1-\Delta)}{a^d}.
\end{align}
In this perspective, the critical point is reached when this energy scale becomes comparable to the typical interaction energy $J_{\mathrm{typ}}$ between spins in the diluted system, leading to the condition
\begin{align}
    \frac{(1-p)(1-\Delta_c)}{a^d}\sim J_\mathrm{typ}\,.
\end{align}
To estimate $J_{\mathrm{typ}}$, one has to consider the typical interaction between spins separated by distances larger than a characteristic distance $r_{\mathrm{typ}}$, weighted by the spin density. In a continuum approximation, this yields
\begin{align}
    J_\mathrm{typ}\sim \frac{(1-p)}{a^d}\int_{r_\mathrm{typ}}^\infty d^drr^{-\alpha} = \frac{(1-p)\Omega_d}{a^d}\frac{r_\mathrm{typ}^{d-\alpha}}{(\alpha-d)}\,.
\end{align}
The typical distance $r_{\mathrm{typ}}$ is obtained by first computing the probability that a spherical region of volume $A=\Omega_d r^d$, with $\Omega_d = \pi^{d/2}/\Gamma(d/2+1)$, contains no spins. Dividing the region into $n$ subcells and taking the limit $n\to\infty$, the probability that the region is empty is
\begin{align}
P_{\mathrm{empty}}(A)
&= \lim_{n\to\infty}
\left(1-\frac{(1-p)A}{a^d n}\right)^n
\notag\\
&= \exp\left[-\frac{(1-p)A}{a^d}\right].
\end{align}
The probability distribution of nearest-neighbor distances is then
\begin{align}
\mathcal{P}(r)
&= \frac{d}{dr}\left[1-P_{\mathrm{empty}}\left(\Omega_d(r^d-a^d)\right)\right] \notag\\
&= \frac{d\Omega_d(1-p)}{a^d}
r^{d-1}
\exp\left[-\frac{(1-p)\Omega_d(r^d-a^d)}{a^d}\right].
\end{align}
From this distribution, the typical distance can be estimated as the geometric mean
\begin{align}
r_{\mathrm{typ}}
\sim e^{\langle\log r\rangle}
\propto \left[\frac{(1-p)\Omega_d}{a^d}\right]^{-1/d}\,,
\end{align}
which implies the scaling
\begin{align}
r_{\mathrm{typ}}\sim (1-p)^{-1/d}\,.
\end{align}

Substituting this result into the expression for $J_{\mathrm{typ}}$, we obtain
\begin{align}
J_{\mathrm{typ}}
\sim (1-p)r_\mathrm{typ}^{d-\alpha}
\sim (1-p)^{\alpha/d}\,.
\end{align}
Finally, equating the two energy scales yields the scaling of the critical anisotropy,
\begin{align}
1-\Delta_c
\sim (1-p)^{\alpha/d-1}
= (1-p)^{2/d_s},
\end{align}
where in the last equality we used the relation $d_s=2d/(\alpha-d)$ for long-range lattices. This result coincides with the scaling obtained from the spectral-gap argument near the percolation transition specialized to long-range lattices in Eq.~\eqref{eq: Deltac p LR diluted}.

\end{document}